\documentclass[aps,prd,preprint,groupedaddress]{revtex4}
\usepackage{graphicx}
\newcommand{\no}{\nonumber}
\newcommand{\sla}[1]{#1\hspace{-5pt}/}
\newcommand{\slal}[1]{#1\hspace{-10pt}/~}

\newcommand{\p}{\mbox{\boldmath $p$}}

\newcommand{\Op}{\cal O}
\newcommand{\la}{\langle}
\newcommand{\ra}{\rangle}
\newcommand{\slabar}[1]{\mbox{$\overline{#1\hspace{-5pt}/}~$}}
\def\ovec{\partial_\mu\hspace{-0.4cm}\raisebox{1.8ex}{$\rightarrow$}}
\def\antivec{\partial_\mu\hspace{-0.4cm}\raisebox{1.8ex}{$\leftarrow$}}
\newcommand{\latt}{{\rm latt}}
\newcommand{\cont}{{\rm cont}}
\newcommand{\lo}{{(0)}}
\newcommand{\mph}{{m_{Q}}}
\newcommand{\mpl}{{m_{q}}}
\newcommand{\mphlo}{\mph^{(0)}}
\newcommand{\mpllo}{\mpl^{(0)}}

\newcommand{\mphlomf}{{\tilde m}_Q^{(0)}}
\newcommand{\pos}{{p^*}}

\newcommand{\qoss}{{q^*_s}}
\newcommand{\qosd}{{q^*_d}}

\newcommand{\msbar}{{\overline {\rm MS}}}
\newcommand{\shq}{\mbox{$\sinh(m_q^\lo)$}}
\newcommand{\chq}{\mbox{$\cosh(m_q^\lo)$}}

\begin{document}

% \preprint{
% RBRC-???
%  }

 \title{
  Perturbative determination of mass dependent renormalization
  and improvement
  coefficients for the heavy-light vector and axial-vector currents with
  relativistic heavy and domain-wall light quarks
 }
 \author{Norikazu Yamada$^a$, Sinya Aoki$^{a,b}$ and Yoshinobu Kuramashi$^b$}
 \affiliation{
 $^a$RIKEN BNL Research Center, Brookhaven National Laboratory,
     Upton, NY 11973, USA\\
 $^b$Institute of Physics, University of Tsukuba,
     Tsukuba, Ibaraki 305-8571, Japan
 }
 \date{\today}
 \begin{abstract}
  We determine the mass dependent renormalization
  as well as improvement
  coefficients for the
  heavy-light vector and axial-vector currents consisting of the
  relativistic heavy and the domain-wall light quarks through the
  standard matching procedure.
  The calculation is carried out perturbatively at the one loop level to
  remove the systematic error of $O(\alpha_s (am_Q)^n a\p)$ as well as
  $O(\alpha_s (am_Q)^n)$ ($n\ge$0), where $\p$ is a typical momentum
  scale in the heavy-light system.
  We point out that
  renormalization and improvement coefficients of the heavy-light
  vector current agree with those of the axial-vector current,
  thanks to the exact chiral symmetry for the light quark.
  The results obtained with three different gauge actions,
  plaquette, Iwasaki and DBW2,
  are presented
  as a function of heavy quark mass and domain-wall height.
 \end{abstract}

 \maketitle

 \section{Introduction}
 \label{sec:intro}

 The study of heavy-light hadrons including a single heavy quark ($b$ or
 $c$) has played a central role in the particle physics.
 In particular, much attentions have been paid to estimating weak decay
 matrix elements of these hadrons to determine the
 Cabbibo-Kobayashi-Maskawa matrix elements.
 Lattice QCD simulation offers a promising method to calculate the
 decay amplitudes from the first principles, and a number of efforts
 using lattice QCD have been already made and contributed to the
 heavy-light physics~\cite{Yamada:2002wh,Kronfeld:2003sd}.
 
 However, the lattice calculation of the heavy-light system is not yet
 precise enough.
 Since the currently accessible inverse lattice spacing is at most
 $1/a\sim$3, 4 GeV, so that $am_{b,c}\ll$1 never holds, the lattice
 calculations suffer from sizable $O((am_{b,c})^n)$ ($n>$1) errors.
 Another difficulty, though this is a rather general problem in Lattice
 QCD simulations of hadrons, arises from chiral extrapolations of
 physical quantities to their physical values ($\sim m_{u,d}$) as
 numerically investigated in
 Refs.~\cite{Aoki:2003xb,Aoki:2001yr,Yamada:2001xp}.
 Because of limited computational resources, we are forced to simulate
 light quarks much heavier than the physical light quark masses, and,
 then, long extrapolations are required.
 In addition, non-analytic behaviors of physical quantities around the
 massless point which chiral perturbation theory predicts make reliable
 extrapolations even more difficult.
 
 One possible way to overcome the above obstacles is to combine the
 relativistic heavy quark~\cite{Aoki:2001ra} and the domain-wall
 light quark~\cite{Kaplan:1992bt,Shamir:1993zy,Furman:ky} to simulate
 heavy-light quark systems.
 The relativistic heavy quark~\cite{Aoki:2001ra} is obtained by pushing
 the on-shell improvement program~\cite{Luscher:1984xn} to the Wilson
 type massive fermions, and its action is equal to the Fermilab
 action~\cite{El-Khadra:1996mp} with a special choice of parameters.
 The relativistic heavy quark action has four parameters to be tuned,
 and the perturbative determination of these parameters as well as the
 wave function renormalization has been completed to the one-loop level
 so far~\cite{Aoki:2003dg}.
 As for the problem of chiral extrapolations in terms of light quark masses,
 we employ the domain-wall formalism for light quarks,
 which has already been used successfully in large scale simulations.
 It has been demonstrated that this formalism
 allows to simulate lighter quarks than the standard Wilson type quark
 action.

 In this paper, we determine the renormalization coefficients for the
 heavy-light vector and axial-vector currents consisting of the
 relativistic heavy and the domain-wall light quarks through the
 standard one-loop matching.
 The improvement coefficients associated with the dimension four operators
 are evaluated to the one-loop as well.
 Since the obtained coefficients depend on both heavy quark mass $m_Q$
 and domain-wall height $M_5$, we work out the interpolation formula to
 express the results as simple as possible.
 In this work we take three different gauge actions, the plaquette, the
 Iwasaki~\cite{Iwasaki:1983ck} and
 DBW2~\cite{Takaishi:1996xj,deForcrand:1999bi}, as examples. 
 The Feynman rules and useful formula for the gauge actions are
 described in Ref.~\cite{Aoki:2003dg} in rather general manner, so we
 will omit that part of explanations from this paper.
 With the coefficients obtained here and the lattice calculations of all
 relevant operators, we can remove
 the systematic error of $O(\alpha_s (am_Q)^n a\p)$ as well as
 $O(\alpha_s (am_Q)^n)$ ($n\ge$0) from physical amplitudes.
 As for the calculational method,
 we completely follows that in Ref.~\cite{Aoki:2004th}, where the
 similar calculations are performed using the clover action for
 the light quark.

 This paper is organized as follows.
 In Sec.~\ref{sec:actions}, we briefly explain the quark actions used in
 the following sections.
 In Sec.\ref{sec:definition-imp-coef}, the definitions of the renormalization
 and improvement
 coefficients which we determine in this paper are given.
 Then, we give a short notice on the benefit of chiral symmetry in
 Sec.\ref{sec:chiral-symmetry}.
 The calculational method and the numerical results are given in
 Secs.~\ref{sec:determination}.
 In Sec.~\ref{sec:imp-of-ptseries}, we consider two ways to improve the
 convergence of the perturbative series.
 Concluding remarks are stated in Sec.~\ref{sec:conlcusion}.
 In the process of this calculation, we found that the on-shell
 improvement of the massive domain-wall fermions is necessary.
 The improvement procedure in this case is briefly explained in
 appendix~\ref{sec:on-shell-imp-dwf}.

 \section{the quark actions}
 \label{sec:actions}
 \subsection{The relativistic heavy quark}
 \label{subsec:heavy}

 In this work, the heavy quark is described by the relativistic heavy quark
 action~\cite{Aoki:2001ra},
 \begin{eqnarray}
    S_Q
= \sum_x&&\!\!\!
          \left[\  m_0{\bar Q}(x)Q(x) +{\bar Q}(x)\gamma_0 D_0Q(x)
                 + \nu_Q \sum_i {\bar Q}(x)\gamma_i D_i Q(x)\right.\no\\
        &&\left. - \frac{r_t a}{2} {\bar Q}(x)D_0^2 Q(x)
                 - \frac{r_s a}{2} \sum_i {\bar Q}(x)D_i^2 Q(x)\right.\no\\
        &&\left. - \frac{ig a}{2}c_E
                   \sum_i     {\bar Q}(x)\sigma_{0i}F_{0i} Q(x)
                 - \frac{ig a}{4}c_B
                   \sum_{i,j} {\bar Q}(x)\sigma_{ij}F_{ij} Q(x)\
          \right].
\label{eq:action_clover}
 \end{eqnarray}
 The parameters appearing in the action are determined by requiring
 $n$-point amputated on-shell Green's function on the lattice to be
 equal to the continuum one up to $O(a^2)$.
 In order to obtain the improvement coefficients which are valid at an
 arbitrary quark mass, no expansion in $(am_0)$ nor $1/(am_0)$ is
 allowed through the calculation.
 By matching the two-point function (propagator) to the continuum one at
 tree level, the pole mass, wave function, $\nu_Q$ and $r_s$ are found
 to be
 \begin{eqnarray}
  m_Q^\lo &=& \ln\left|\frac{m_0+r_t+\sqrt{m_0^2+2r_t m_0+1}}
                        {1+r_t}\right|,
  \label{eq:polemassQ}\\
  Z_{Q,\latt}^\lo &=& {\rm cosh}(m_Q^\lo)+r_t\,{\rm sinh}(m_Q^\lo),
  \label{eq:ZQlo}\\
  \nu_Q^\lo &=& \frac{{\rm sinh}(m_Q^\lo)}{m_Q^\lo},
  \label{eq:nuQ}\\
  r_s^\lo &=& \frac{{\rm cosh}(m_Q^\lo)+r_t\,{\rm sinh}(m_Q^\lo)}{m_Q^\lo}
             -\frac{{\rm sinh}(m_Q^\lo)}{\left(m_Q^\lo\right)^2}
           =  \frac{1}{m_Q^\lo}(Z_Q^\lo - \nu_Q^\lo).
  \label{eq:rsQ}
 \end{eqnarray}
 Matching the amputated three-point function (quark-gluon form factor)
 leads to
 \begin{eqnarray}
  c_E^\lo &=& r_t\,\nu_Q^\lo,
  \label{eq:ceQ}\\
  c_B^\lo &=& r_s^\lo
  \label{eq:cbQ}.
 \end{eqnarray}
Since $r_t$ is not constrained from the improvement condition, in the
following we set $r_t$=1 for simplicity.
Once the parameters have been tuned to their proper values given in
eqs.~(\ref{eq:polemassQ})--(\ref{eq:cbQ}), we can remove the systematic
errors of $O((am_Q)^n)$ and $O((am_Q)^n a\p)$ ($n\ge$0) from the action.

Extending the above procedure to the one-loop level,
$O(\alpha_s(am_Q)^n)$ and $O(\alpha_s (am_Q)^n a\p)$ ($n\ge$0) errors
are removed.
This calculation was carried out in Ref.~\cite{Aoki:2003dg}, and the
numerical results depending on the quark mass are given in terms of an
interpolation formula.
Here we only quote the result of the wave function renormalization for
the later use.
Defining the wave function renormalization
by
\begin{eqnarray}
     Q_{\msbar}
 &=& \sqrt{Z_Q(m_Q^\lo)}\,Q_\latt,\\
     Z_Q(m_Q^\lo)
 &=& \frac{Z_{Q,\latt}(m_Q^\lo)}{Z_{Q,\cont}(m_Q^\lo)}
  =  Z_{Q,\latt}^\lo(m_Q^\lo)\,
     \left[ 1 + g^2 \bigg( - \frac{C_F}{16\pi^2}
      \ln(a^2\mu^2)
                           + \Delta_Q(m_Q^\lo)
                    \bigg)
     \right],
\label{eq:z_akt-1-loop}
\end{eqnarray}
$\Delta_Q(m_Q^\lo)$ is, to good precision, given by
\begin{eqnarray}
     \Delta_Q(m_Q^\lo)
 =   \Delta_Q(0)
   + \frac{  \sum_{i=1}^4 a_i\{m_Q^\lo\}^i}
          {1+\sum_{i=1}^4 b_i\{m_Q^\lo\}^i},
\end{eqnarray}
where the values of $\Delta_Q(0)$, $a_i$ and $b_i$
($i=1,\dots,4$) are listed in Table II of Ref.~\cite{Aoki:2003dg}.

It should be noted that in the massless limit all $\nu_Q^{(0)}$,
$r_s^{(0)}$, $c_E^{(0)}$ and $c_B^{(0)}$ are reduced to unity
(for $r_t$=1), and the action is reduced to the $O(a)$ improved SW quark
action~\cite{Sheikholeslami:1985ij}.
More importantly, the space-time rotational symmetry is restored in this
limit to all orders of $a$ and $\alpha_s$.

 \subsection{The domain-wall quark action}
 \label{subsec:light}

 As for light quarks, we take the domain-wall
 fermion~\cite{Kaplan:1992bt,Shamir:1993zy,Furman:ky}.
 More precisely, we employ the following version of the domain-wall
 action in this work,
 \begin{eqnarray} 
     S_{\rm DW}
 &=& \sum_{x,y}\sum_{s,t}
     \bar{\psi}_s(x)D_{\rm DW}(x,s;y,t)\psi_t(y),
\label{eq:action_dw}\\
       D_{\rm DW}(x,s;y,t)
 &=&   D_{\rm W}(x,y)\,\delta_{s,t}
     + \delta_{x,y}D_5(s,t),\\
     D_{\rm W}(x,y)
 &=&   \sum_\mu\gamma_\mu D_\mu - \frac{a}{2} \sum_\mu D_\mu^2
     - M_5\,\delta_{x,y},
\label{eq:wilson-dirac-op}\\
       D_5(s,t)
 &=&   \delta_{s,t} - P_L\,\delta_{s+1,t} - P_R\,\delta_{s-1,t}
     + m_f\left[   P_L\,\delta_{s,N_5}\delta_{1,t}
                 + P_R\,\delta_{s,1}  \delta_{N_5,t}
          \right],
  \end{eqnarray}
 where $P_{R/L}=(1\pm\gamma_5)/2$, and $s$ and $t$ label the coordinate
 of the fifth dimension running 1 to $N_5$.
 We set $N_5\rightarrow\infty$ throughout this paper.
 The physical quark field $q(x)$ is, then, constructed by
 \begin{eqnarray}
      q(x)
  &=& P_L\psi_1(x) + P_R\psi_{N_5}(x),\\
      \bar{q}(x)
  &=& \bar{\psi}_1(x)P_R + \bar{\psi}_{N_5}(x)P_L.
 \end{eqnarray}
 As long as massless quarks are concerned, this original action is
 enough to proceed.
 However, as explained in Ref.~\cite{Aoki:2004th} and later as well, in
 a part of calculations we should keep the light quark mass small but
 finite.
 Once quark masses become finite, the infrared divergences which do not
 exist in the continuum show up in a certain class of loop integrations
 due to the mismatch of dispersion relation and make perturbative
 matching to $O(\alpha_sa\p)$ meaningless.
 To eliminate the mismatch, the original domain-wall action is required
 to be improved, and the improvement can be realized by simply applying
 the on-shell improvement program~\cite{Luscher:1984xn,Aoki:2001ra}
 to the massive domain-wall fermion.
 As a consequence, it is found that we need, at least, two modifications
 in the original domain-wall action.
 First we replace the first term of eq.~(\ref{eq:wilson-dirac-op}) with
 \begin{eqnarray}
 \gamma_0 D_0 + \nu_q \sum_i\gamma_iD_i,
  \label{eq:newterm1}
 \end{eqnarray}
 and, then, add a new term,
 \begin{eqnarray}
   - \frac{R_s}{2}\ \bar{q}(x)\ \sum_iD_i^2\ q(x)
 = - \frac{R_s}{2}\ \bar{\psi}_s(x)\
          \sum_iD_i^2\
          \left(   P_L\ \delta_{s,N_5}\delta_{t,1}
                 + P_R\ \delta_{s,1}\delta_{t,N_5}
          \right)\ \psi_t(x),
  \label{eq:newterm2}
 \end{eqnarray}
 to the action eq.~(\ref{eq:action_dw}).
 Here we introduced two parameters $\nu_q$ and $R_s$, which can be tuned
 by following the on-shell improvement program~\cite{Aoki:2001ra}.
 As a result (see appendix~\ref{sec:on-shell-imp-dwf} for the
 derivations), we obtain
 \begin{eqnarray}
  \nu_q^\lo(m_q^\lo) &=& \frac{\shq}{m_q^\lo},
   \label{eq:nuq}\\
     R_s^\lo(m_q^\lo,M_5)
 &=& \frac{m_f}{m_q^\lo\,\shq\,(1-m_f^2)}\no\\
 & & \times\Bigg[\
       (1+m_f^2)\left(\chq-\frac{\shq}{m_q^\lo}\right)
     - 2\,m_f\,(2-M_5)\big(\, \shq-m_q^\lo \big) \no\\
 & &\hspace*{4ex}
     - 2\,m_f\,m_q^\lo
       \left(\chq-\frac{(\shq)^2}{(m_q^\lo)^2}\right)\
           \Bigg],
   \label{eq:R_s-tune}
 \end{eqnarray}
 where $m_q^\lo$ is the tree-level pole mass depending on $m_f$ and $M_5$,
 and its expression is given in eq.~(\ref{eq:polemass}).
 As $m_f\rightarrow 0$, these parameters approach to
 \begin{eqnarray}
  \nu_q^\lo &\rightarrow& 1 + O(m_f^2),\\
  R_s^\lo   &\rightarrow& \frac{m_f}{3} + O(m_f^3),
 \end{eqnarray}
 respectively, and, thus, the action is reduced to the original one.
 Analytical studies of the loop integrations reveal that the tree level
 matching of $\nu_q$ and $R_s$ suffices to get rid of the extra
 infrared divergences associated with massive light quarks,
 so that it is not necessary to introduce terms like the clover terms in this
 work though to remove the $O((am_q^\lo)^na\p)$ errors from the action
 throughly we still need the clover terms whose tree-level coefficients are
 calculated in appendix~\ref{sec:on-shell-imp-dwf}.

 For the wave function renormalization, the one for the massless
 domain-wall fermions is sufficient for the following calculation.
 The relation of the massless domain-wall quark field to the continuum
 one with $\msbar$ in NDR is given by
 \begin{eqnarray}
     q_{\msbar}
 &=& \sqrt{\frac{Z_q(w_0)}{(1-w_0^2)\,Z_w(w_0)}}\,q_{\latt},\\
     w_0
 &=& 1-M_5,\\
     Z_{q}(w_0)
 &=& 1 + g^2
     \left( - \frac{C_F}{16\pi^2}
      \ln(a^2\mu^2)
            + \Delta_q(w_0)
     \right),
\label{eq:z_dwf-1-loop}\\
     Z_w(w_0)
 &=& 1 + g^2 \Delta_w(w_0),
 \end{eqnarray}
 and the numerical values of $\Delta_q(w_0)$ and $\Delta_w(w_0)$ for
 various values of $w_0$ ($M_5$) are constructed from $z_2({\rm DRED})$
 and $\Sigma_w$, which are tabulated in Tables I, III and V of
 Ref.~\cite{Aoki:2002iq} for the plaquette, Iwasaki and DBW2 gauge
 actions respectively, through the relations,
 \begin{eqnarray}
     \Delta_q(w_0)
 &=& \frac{C_F}{16\pi^2}z_2({\rm NDR})
  =  \frac{C_F}{16\pi^2}\,\left(z_2({\rm DRED})+1\right),\\
     \Delta_w(w_0)
 &=& \frac{C_F}{16\pi^2}\,\frac{2w_0}{1-w_0^2}\Sigma_w.
 \end{eqnarray}
 The space-time rotational symmetry is recovered in the massless limit
 to all orders of $a$ and $\alpha_s$.

 \section{definition of the improvement coefficients}
 \label{sec:definition-imp-coef}

 Taking into account existing symmetries such as parity, charge
 conjugation, {\it etc.}, it is turned out that the renormalized
 continuum vector and axial-vector currents can be expressed in terms of
 $O(a)$ improved lattice operators as
 \begin{eqnarray}
     V^{\msbar}_\mu(x)
   = Z_{V_\mu}^{\msbar-\latt}
   &&\!\!\!\!\left[\,
              {\bar q(x)} \gamma_\mu Q(x)
            - g^2 c_{V_\mu}^+\partial_\mu^- \{{\bar q(x)} Q(x)\}
            - g^2 c_{V_\mu}^- \partial_\mu^+ \{{\bar q(x)} Q(x)\}\right.\no\\
   && \left.- g^2 c_{V_\mu}^L \{{\vec\partial_i}{\bar q(x)}\}
              \gamma_i\gamma_\mu Q(x) 
            - g^2 c_{V_\mu}^H {\bar q(x)}\gamma_\mu \gamma_i
              \{{\vec \partial_i} Q(x)\}+O(g^4)\right],
 \label{eq:v_r}\\
    A^{\msbar}_\mu(x)
  = Z_{A_\mu}^{\msbar-\latt}
    &&\!\!\!\!\left[\,
            {\bar q(x)} \gamma_\mu\gamma_5 Q(x)
          - g^2 c_{A_\mu}^+\partial_\mu^+
            \{{\bar q(x)} \gamma_5 Q(x)\}
          - g^2 c_{A_\mu}^- \partial_\mu^-
            \{{\bar q(x)}\gamma_5 Q(x)\}\right.\no\\
&&  \left.+ g^2 c_{A_\mu}^L \{{\vec\partial_i}{\bar q(x)}\}
            \gamma_i \gamma_\mu\gamma_5 Q(x) 
          - g^2 c_{A_\mu}^H {\bar q(x)}\gamma_\mu\gamma_5 \gamma_i
            \{{\vec \partial_i} Q(x)\}+O(g^4)
    \right].
\label{eq:a_r}
 \end{eqnarray}
 where $\partial^\pm$=$\ovec\pm\antivec$, $q(x)$ and $Q(x)$ represent
 light and heavy quark fields defined on the lattice, respectively, and
 the coefficients
 $Z_{V_\mu,A_\mu}^{\msbar-\latt}$ and
 $c_{V_\mu,A_\mu}^{\pm,H,L}$ are the ones we determine in the
 following sections.
 While the right hand side can be written in a gauge invariant way by
 replacing partial derivatives with covariant derivatives, the above
 expression suffices for the following discussion.
 Although the general form given in eqs.~(\ref{eq:v_r}) and
 (\ref{eq:a_r}) does not depend on details of the lattice actions,
 the coefficients
 $Z_{V_\mu,A_\mu}^{\msbar-\latt}$ and
 $c_{V_\mu,A_\mu}^{\pm,H,L}$ do.
 Since the space-time rotational symmetry is, in general, violating on the
 lattice unless quarks are massless, the fourth and fifth terms on the
 right hand side are required to improve the violation.
 Therefore, although the equation of motion always allows us to set
 $c^H_{V_0,A_0}=c^L_{V_0,A_0}=0$ and we take this, it never means that
 the space-time rotational symmetry is recovered in the temporal
 component of the currents.

 \section{benefit of chiral symmetry}
 \label{sec:chiral-symmetry}

 Here let us mention an advantage of using lattice chiral fermions to
 describe the light quark in heavy-light system by taking the
 heavy-light vector and axial-vector currents as examples.
 The similar discussion is made in Ref.~\cite{Becirevic:2003hd}, where
 the renormalization of the static-light current is calculated using the
 overlap fermion for the light quark.
 The authors of Ref.~\cite{Gulez:2003uf} also mentioned the similar
 point in the calculation with the NRQCD heavy and the naive light
 quarks.

 Now let us consider a system including $N_f$=2
 massless light quarks $q^{i}(x)$ ($i=1,2$), described by
 the lattice fermions with the exact chiral symmetry, and
 one heavy quark $Q(x)$.
 An infinitesimal flavor non-singlet axial transformation is defined by
 \begin{eqnarray}
     \delta_A\, q^i(x)
 &=& i w^a(x)\,\frac{(\tau^a)^{ij}}{2}\,\gamma_5\,q^j(x),\\
     \delta_A\, \bar q^i(x)
 &=& \bar q^j(x) i w^a(x)\,\frac{(\tau^a)^{ji}}{2}\,\gamma_5,
   \label{eq:chiral-rotaion}
 \end{eqnarray}
 where $i$ and $j$ are the flavor indexes and run 1 to $N_f$, and
 $\tau^a$ is the Pauli matrix.
 Hereafter we take an arbitrary transformation parameter $w^a(x)$ as
 \begin{eqnarray}
  w^a(x) = \left\{\begin{array}[c]{ll}
            \delta_{a,3} & \mbox{for $x$ $\in$ R}\\
            0   & \mbox{otherwise}
                  \end{array}
           \right.,
 \end{eqnarray}
 where $R$ is a certain space-time region with smooth boundary
 $\partial R$.
 The exact chiral symmetry for the light quarks ensures the existence
 of the conserved axial-vector current.
 Using this fact, we can derive the following axial Ward-Takahashi
 identity on the lattice
 \begin{eqnarray}
    \la 0|\ i \int_{\partial R}\!\! d\sigma_\mu(y)\,
          A^{3,\rm ll}_\mu(y)\,{\Op}_{\rm int}\,{\Op}_{\rm ext}\ |0\ra
  = \la 0|\ (\delta_A {\Op}_{\rm int})\,{\Op}_{\rm ext}\ |0\ra,
 \label{eq:pcac-general}
 \end{eqnarray}
 where $A^{3,\rm ll}_\mu(y)$ is the third component of the lattice
 conserved light-light flavor non-singlet axial-vector current, and
 ${\Op}_{\rm int}$ and ${\Op}_{\rm ext}$ are arbitrary operators
 localized in the interior and exterior of $R$, respectively.
 Now suppose
 \begin{eqnarray}
  {\Op}_{\rm int}=V_\nu^{i, \rm hl}(x)=\bar{q}^i(x)\gamma_\nu Q(x),
 \end{eqnarray}
 then,
 \begin{eqnarray}
    \delta_A{\Op}_{\rm int}
  = - i\,\bar{q}^j(x)\frac{1}{2}(\tau^3)^{ji}\gamma_\nu\gamma_5 Q(x)
  = - \frac{i}{2}(\tau^3)^{ji} A_\nu^{j,\rm hl}(x).
      \label{eq:delta_A_V}
 \end{eqnarray}
 is read.
 Substituting the above into eq.~(\ref{eq:pcac-general}), one would reach
 \begin{eqnarray}
   \la 0|\,\int_{\partial R}d\sigma_\mu(y)
    A^{3,\rm ll}_\mu(y)\,V_\nu^{1,\rm hl}(x)\,
    {\Op}_{\rm ext}\ |0\ra
  = - \frac{1}{2}\,\la 0|\ A_\nu^{1,\rm hl}(x)\,
      {\Op}_{\rm ext}\ |0\ra,
 \label{eq:pcac-hl}
 \end{eqnarray}
 for $i=1$.

 Now repeat the above discussion with the renormalized currents,
 and rewriting them in terms of the bare currents, one obtains
 \begin{eqnarray}
    Z_{V_\nu}^{\msbar-\latt}
   \la 0|\,\int_{\partial R}d\sigma_\mu(y)
    A^{3,\rm ll}_\mu(y)\,V_\nu^{1,\rm hl}(x)\,
    {\Op}_{\rm ext}\ |0\ra
  = - Z_{A_\nu}^{\msbar-\latt}
      \frac{1}{2}\,\la 0|\ A_\nu^{1,\rm hl}(x)\,
      {\Op}_{\rm ext}\ |0\ra,
 \label{eq:renormalized-pcac-hl}
 \end{eqnarray}
 Here it is important to note that the conserved current does not
 receive the renormalization.
 Comparing eq.~(\ref{eq:pcac-hl}) to
 eq.~(\ref{eq:renormalized-pcac-hl}), the relation
 \begin{eqnarray}
      Z_{V_\mu}^{\msbar-\latt}
  &=& Z_{A_\mu}^{\msbar-\latt}
  \label{eq:chiral-relation-start} 
 \end{eqnarray}
 is obtained.
 The operator ${\Op}_{\rm int}$ is arbitrary, so that repeating the
 above procedure with the right hand side of eq.(\ref{eq:v_r}) and
 comparing the resultant expression with eq.~(\ref{eq:a_r}),
 \begin{eqnarray}
  c_{V_\mu}^+ &=& - c_{A_\mu}^-,\\
  c_{V_\mu}^- &=& - c_{A_\mu}^+,\\
  c_{V_k}^H   &=&   c_{A_k}^H,\\
  c_{V_k}^L   &=&   c_{A_k}^L,
  \label{eq:chiral-relation-end}
 \end{eqnarray}
 are derived in addition to eq.~(\ref{eq:chiral-relation-start}).
 Thus it is proved that, whatever the heavy quark action is, if a
 light quark field on the lattice has chiral symmetry, the
 renormalization constants and the improvement coefficients for vector
 and axial-vector currents are related to each other.
 This statement can be extended to any pair of operators which belong to
 the same chiral multiplet.

 \section{One-loop determination of the coefficients}
 \label{sec:determination}

 \subsection{Calculational method}
 \label{subsec:method}

\begin{figure}[t]
 \begin{center}
  \includegraphics*[width=8cm,clip=true]{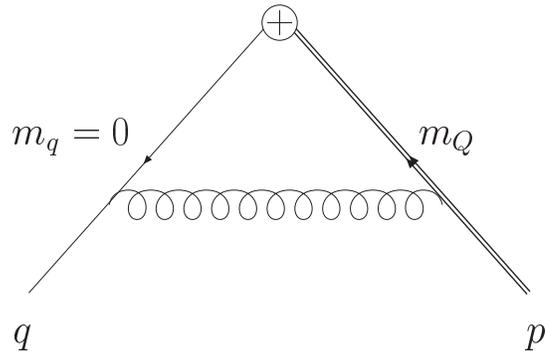}
 \end{center}
 \caption{One-loop diagrams for the vertex functions. $q$ denotes the
          outgoing momentum and $p$ denotes the incoming momentum.}
 \label{fig:vtx_1loop}
\end{figure}

 In this section, the calculational method is outlined by taking the
 vector current as an example.
 More complete explanations are found in Ref.~\cite{Aoki:2004th}, where
 the similar calculation is performed with the clover light quark.
 Let us first define the off-shell vertex function of the lattice vector
 current $\bar{q}(x)\gamma_\mu Q(x)$ by
 \begin{eqnarray}
    \Lambda_\mu^\latt(p,q,\mph,\mpl,M_5)
  = \sum_{i=0} (g^2)^i \Lambda_\mu^{(i),\latt}(p,q,\mph,\mpl,M_5).
 \end{eqnarray}
 The tree level contribution $\Lambda^{(0)}_\mu$=$\gamma_\mu$ is
 immediately followed.
 The one-loop contribution $\Lambda_\mu^{(1)}(p,q,\mph,\mpl,M_5)$,
 obtained by calculating Fig.~\ref{fig:vtx_1loop}, can be
 written in the following general form,
 \begin{eqnarray}
      \Lambda_\mu^{(1),\latt}(p,q,\mph,\mpl,M_5)
&=&   \gamma_\mu F_1^\mu
    + \gamma_\mu\{{\sla{p}} F_2^\mu + \slal{p_s} F_3^\mu\}
    + \{{\sla{q}} F_4^\mu+\slal{q_s} F_5^\mu\}\gamma_\mu\no\\
 && + {\sla{q}}\gamma_\mu\,\sla{p} F_6^\mu
    + {\sla{q}}\gamma_\mu\slal{p_s} F_7^\mu
    + {\slal{q_s}}\gamma_\mu\sla{p} F_8^\mu
    + \gamma_\mu\slal{p_s}\sla{p} F_9^\mu
    + {\sla{q}}\slal{q_s}\gamma_\mu F_{10}^\mu \no\\
 && + (p_\mu+q_\mu)\left[ G_1^\mu+\sla{p} G_2^\mu
    + \sla{q} G_3^\mu+\sla{q}\sla{p} G_4^\mu\right] \no\\
 && + (p_\mu-q_\mu)\left[ H_1^\mu+\sla{p} H_2^\mu
    + \sla{q} H_3^\mu+\sla{q}\sla{p} H_4^\mu\right]
    + O(a^2),
\label{eq:v_offsh}
 \end{eqnarray}
where
\begin{eqnarray}
\sla{p}   =\sum_{\alpha=0}^3 p_\alpha \gamma_\alpha,\ \
\sla{q}   =\sum_{\alpha=0}^3 q_\alpha \gamma_\alpha,\ \
\slal{p_s}=\sum_{i=1}^3 p_i \gamma_i,\ \
\slal{q_s}=\sum_{i=1}^3 q_i \gamma_i,
\end{eqnarray}
and the coefficients $F_I^\mu$, $G_I^\mu$ and $H_I^\mu$ are the scalar
functions of $p^2$, $q^2$, $p\cdot q$, $\mph$, $\mpl$ and $M_5$.
Here it should be noted that in $\mu$=0 case we can set $F_I^0$=0 for
$I$=3, 5, 7, 8, 9 and 10 without loss of generality.

Sandwiching eq.~(\ref{eq:v_offsh}) by the on-shell quark states $u_Q(p)$
and $\bar u_q(q)$, which satisfy $\sla{p} u_Q(p) = i \mph u_Q(p)$ and
$\bar u_q(q) \sla{q} = i\mpl \bar u_q(q)$, 
the matrix elements are reduced to 
\begin{eqnarray}
& & {\bar u}_q(q)\Lambda_\mu^{(1),\latt}(p,q,\mph,\mpl,M_5)u_Q(p)\no\\
=&& X_\mu^{\latt} {\bar u}_q(q)\gamma_\mu u_Q(p)
  + R_\mu^{\latt} {\bar u}_q(q)\gamma_\mu\,\slal{p_s} u_Q(p)
  + S_\mu^{\latt} {\bar u}_q(q)\,\slal{q_s}\gamma_\mu u_Q(p)
   \no\\
&+& Y_\mu^{\latt} (p_\mu+q_\mu){\bar u}_q(q)u_Q(p)
  + Z_\mu^{\latt} (p_\mu-q_\mu){\bar u}_q(q)u_Q(p)
  + O(a^2),
\label{eq:v_onsh}
\end{eqnarray}
where we defined
\begin{eqnarray}
X_\mu^\latt&=&F_1^\mu+i\mph F_2^\mu+i\mpl F_4^\mu-\mph\mpl F_6^\mu,
\label{eq:xmu-lat}\\
Y_\mu^\latt&=&G_1^\mu+i\mph G_2^\mu+i\mpl G_3^\mu-\mph\mpl G_4^\mu, \\
Z_\mu^\latt&=&H_1^\mu+i\mph H_2^\mu+i\mpl H_3^\mu-\mph\mpl H_4^\mu, \\
R_k^\latt&=&F_3^k+i\mpl F_7^k+i\mph F_9^k, \\
S_k^\latt&=&F_5^k+i\mph F_8^k+i\mpl F_{10}^k.
\label{eq:c_onsh}
\end{eqnarray}

By repeating the above procedure in the continuum, we obtain the
continuum counterparts, $X^\cont_\mu$, $Y^\cont_\mu$ and $Z^\cont_\mu$.
$R^\cont_k=S^\cont_k=0$ in the continuum because of the presence of the
space-time rotational symmetry.
For $X^\cont_\mu$, the ultraviolet divergence is subtracted by the  
$\msbar$ scheme, and using the resulting expression we define
$X^{\cont,{\rm R}}_\mu$ as
\begin{eqnarray}
      X^{\cont}_\mu
    - \frac{C_F}{16\pi^2}
      \left(\frac{1}{\epsilon}+\ln(4\pi)-\gamma_E
      \right)
  =   X^{\cont,{\rm R}}_\mu
    + \frac{C_F}{16\pi^2}\ln(\mu^2),
\end{eqnarray}
where we factored out the $\ln(\mu^2)$ term
in order to make the $\mu$ dependence clear.
The renormalization factor of the vector currents is then given by
\begin{eqnarray}
    Z_{V_\mu}^{\msbar-\latt}
&=& \sqrt{Z_{Q}(\mphlo)}
    \sqrt{\frac{Z_q(w_0)}{(1-w_0^2)Z_w(w_0)}}
    \left[1-g^2\left( - \frac{C_F}{16\pi^2}\ln(a^2\mu^2)
                      + \Delta_{\gamma_\mu}\right)\right]\no\\
&=& \sqrt{\frac{Z_{Q,\latt}^\lo(\mphlo)}{(1-w_0^2)Z_w(w_0)}}
    \left[1-g^2\Delta_{V_\mu}\right],
\label{eq:z_v}\\
    \Delta_{V_\mu}
&=& \Delta_{\gamma_\mu}
   -\frac{\Delta_{Q}}{2}-\frac{\Delta_{q}}{2},
\end{eqnarray}
where
\begin{eqnarray}
   \Delta_{\gamma_\mu}
 =   X_\mu^\latt - X_\mu^{\cont,{\rm R}}.
  \label{eq:d_v}
\end{eqnarray}
The $\ln(a^2\mu^2)$ term in the first line of eq.~(\ref{eq:z_v}) is
canceled by those in eqs.~(\ref{eq:z_akt-1-loop}) and
(\ref{eq:z_dwf-1-loop}), so the renormalization factor
$Z_{V_\mu}^{\msbar-\latt}$ is independent of the
renormalization scale $\mu$.
Although we take the naive dimensional regularization (NDR) throughout
this paper to regularize ultraviolet divergences, the conversion to the
dimensional reduction (DRED) can be made by
\begin{eqnarray}
   Z_{V_\mu,A_\mu}^{\msbar-\latt}({\rm DRED})
 = Z_{V_\mu,A_\mu}^{\msbar-\latt}({\rm NDR})
 + \frac{1}{2} g^2,
\end{eqnarray}
for both vector and axial-vector currents.

The coefficients $c_{V_\mu}^{\pm,H,L}$ defied in eq.~(\ref{eq:v_r}) are
found to be given by
\begin{eqnarray}
 ic_{V_\mu}^+  &=& Y_\mu^\latt - Y_\mu^\cont,\label{eq:+_v}\\
 i c_{V_\mu}^- &=& Z_\mu^\latt - Z_\mu^\cont,\label{eq:-_v}\\
 -i c_{V_k}^L  &=& S_k^\latt,                \label{eq:l_k_v}\\
 i c_{V_k}^H   &=& R_k^\latt.                \label{eq:h_k_v}
\end{eqnarray}
Since the space-time rotational symmetry is restored at $m_Q=m_q=0$
even on
the lattice, $R_k^\latt|_{m_Q=m_q=0}=S_k^\latt|_{m_Q=m_q=0}=0$, hence,
$c_{V_k}^{H,L}|_{m_Q=m_q=0}=0$.

To calculate the coefficients $\Delta_{\gamma_\mu}$ and
$c_{V_\mu}^{\pm,H,L}$, we define the difference between the lattice
one-loop vertex function and the continuum one by
\begin{eqnarray}
     \lambda_\mu^{(1)}(p,q,m_Q,m_q,M_5)
 =   \Lambda_\mu^{(1),\latt}(p,q,m_Q,m_q,M_5)
   - \Lambda_\mu^{(1),\cont}(p,q,m_Q,m_q,M_5).
\end{eqnarray}
While each of $\Lambda_\mu^{(1),\latt}$ and $\Lambda_\mu^{(1),\cont}$
has infrared divergences, they completely cancel in the difference.
After subtracting the ultraviolet divergences and $\mu$ dependent terms
from $\Lambda_\mu^{(1),\cont}$, $\Delta_{\gamma_\mu}$ and
$c_{V_\mu}^{\pm,H,L}$ are extracted by proper $\gamma$-matrix
projections on $\lambda_\mu^{(1)}$ as follows,
\begin{eqnarray}
     \Delta_{\gamma_k}
 &=& \frac{1}{4}{\rm Tr}
     \left[   \lambda_k^{(1)}(1+\gamma_0)\gamma_k
     \right]_{p=\pos,q=\qosd},
\label{eq:v_k_x}\\
     c_{V_k}^+ + c_{V_k}^-
 &=& \frac{-i}{4}{\rm Tr}
     \left[   \frac{\partial \lambda_k^{(1)}}{\partial p_k}(1+\gamma_0)
            - \frac{\partial \lambda_k^{(1)}}{\partial p_i}(1+\gamma_0)
              \gamma_i\gamma_k
     \right]^{i\ne k}_{p=\pos,q=\qoss},
\label{eq:v_k_ypz}\\
     c_{V_k}^+ - c_{V_k}^-
 &=& \frac{-i}{4}{\rm Tr}
     \left[   \frac{\partial \lambda_k^{(1)}}{\partial q_k}(1+\gamma_0)
            - \frac{\partial \lambda_k^{(1)}}{\partial q_i}(1+\gamma_0)
              \gamma_k\gamma_i
     \right]^{i\ne k}_{p=\pos,q=\qoss},
\label{eq:v_k_ymz}\\
     \Delta_{\gamma_k} - 2\mph\, c_{V_k}^H
 &=& \frac{1}{4}{\rm Tr} 
     \left[   \lambda_k^{(1)}\gamma_k(1+\gamma_0)
            + 2i\mph \frac{\partial \lambda_k^{(1)}}{\partial p_i}
              (1+\gamma_0)\gamma_i\gamma_k
     \right]^{i\ne k}_{p=\pos,q=\qoss},
\label{eq:v_k_r}\\
     \Delta_{\gamma_k} + 2\mpl\, c_{V_k}^L
 &=& \frac{1}{4}{\rm Tr} 
     \left[   \lambda_k^{(1)}(1+\gamma_0)\gamma_k
            + 2i\mpl \frac{\partial \lambda_k^{(1)}}{\partial q_i}
              (1+\gamma_0)\gamma_k\gamma_i
     \right]^{i\ne k}_{p=\pos,q=\qoss},
\label{eq:v_k_s}\\
     \Delta_{\gamma_0}
 &=& \frac{1}{4}{\rm Tr}
     \left[\lambda_0^{(1)}\gamma_0
         - i\mph \frac{\partial \lambda_0^{(1)}}{\partial p_k}
           (1+\gamma_0)\gamma_k
         + i\mpl \frac{\partial \lambda_0^{(1)}}{\partial q_k}
           (1+\gamma_0)\gamma_k
     \right]_{p=\pos,q=\qosd},
\label{eq:v_0_x}\\
 & &\hspace{-18ex}
      \Delta_{\gamma_0}
    - \mph(c_{V_0}^+ + c_{V_0}^-)
    + \mpl(c_{V_0}^+ - c_{V_0}^-)\no\\
 &=&  \frac{1}{4}{\rm Tr}
      \left[   \lambda_0^{(1)}(1+\gamma_0)
             + 2i\mpl \frac{\partial \lambda_0^{(1)}}{\partial q_k}
               (1+\gamma_0)\gamma_k
      \right]_{p=\pos,q=\qosd},
\label{eq:v_0_xyz_d}\\
 & &\hspace{-18ex}
      \Delta_{\gamma_0}
    - \mph(c_{V_0}^+ + c_{V_0}^-)
    - \mpl(c_{V_0}^+ - c_{V_0}^-)
  =   \frac{1}{4}{\rm Tr}
      \left[   \lambda_0^{(1)}(1+\gamma_0)
      \right]_{p=\pos,q=\qoss},
\label{eq:v_0_xyz_s}
\end{eqnarray}
where $i$, $k$=1,2,3 and the external momentum $p$ is set to
$\pos\equiv(im_Q, \vec{0})$ and $q$ to $\qoss\equiv(im_q,\vec{0})$
or $\qosd\equiv(-im_q,\vec{0})$.

All coefficients $\Delta_{\gamma_\mu}$ and $c_{V_\mu}^{\pm,H,L}$ are
obtained from the proper linear combinations of
eqs.~(\ref{eq:v_k_x})-(\ref{eq:v_0_xyz_s}).
As seen from eqs.~(\ref{eq:v_k_s}), (\ref{eq:v_0_xyz_d}) and
(\ref{eq:v_0_xyz_s}), however, $c_{V_k}^L$ and $c_{V_0}^{\pm}$ can not be
extracted at $\mpllo$=0 directly.
Therefore, we keep the light quark mass small but finite in the
calculations of these coefficients, where the on-shell improvement for
the massive domain-wall fermions described in
Sec.~\ref{sec:definition-imp-coef} and
appendix~\ref{sec:on-shell-imp-dwf} is required to remove the associated
infrared divergences.
We estimate these coefficients with two different values of light quark
mass $\mpllo$=0.0001, 0.001, and extrapolated linearly to their massless
limit.

The loop integrations on the lattice are performed by a mode sum for a
periodic box of a size $L^4$ with $L$=48 after transforming the momentum
variable $k_\mu$ to $k'_\mu = k_\mu-\sin k_\mu$.
To search for the $\mph$ and $M_5$ dependences, we take 18 or more
values of $\mph$, depending on $M_5$, in $0.1\le\mph^\lo\le 10.0$ and 11
values of $M_5$ in $0.1\le M_5\le 1.9$.
We choose three different gauge actions, the plaquette, Iwasaki and
DBW2, in this work.
To see the uncertainties, a large part of integrations are
performed with BASES~\cite{Kawabata:1985yt} as well.
For almost all of $m_Q^\lo$ and $M_5$ that we investigated,
all the coefficients using BASES show a nice agreement with those with
the mode sum, while the visible discrepancies appear in
$\Delta_{\gamma_0}$, $c_{V_k}^H$ and $c_{V_k}^{\pm}$
at $M_5$=0.1 and 1.9, and the discrepancies grow as $m_Q$ increases.
Repeating the corresponding integrals using the mode sum with several
values of $L$, it is found that the results approach to the BASES
results as $L$ increases.
From the above observations, we restrict the data within $m_Q^\lo < 7.0$
and $0.1<M_5<1.9$.
After this restriction, the remaining largest discrepancy reduces to
less than 0.001, which can only affect physical amplitudes by 0.5\%, at
most.

The continuum part is evaluated in the naive dimensional regularization
(NDR) with the modified minimal subtraction scheme ($\msbar$).
To make cancellations of infrared divergences explicit suitable
integrands, which can be estimated analytically, are subtracted from the
lattice integrands and added to the continuum ones as done in
Ref.~\cite{Aoki:2004th}.

 \subsection{Numerical results}
 \label{subsec:results}

\begin{figure}[p]
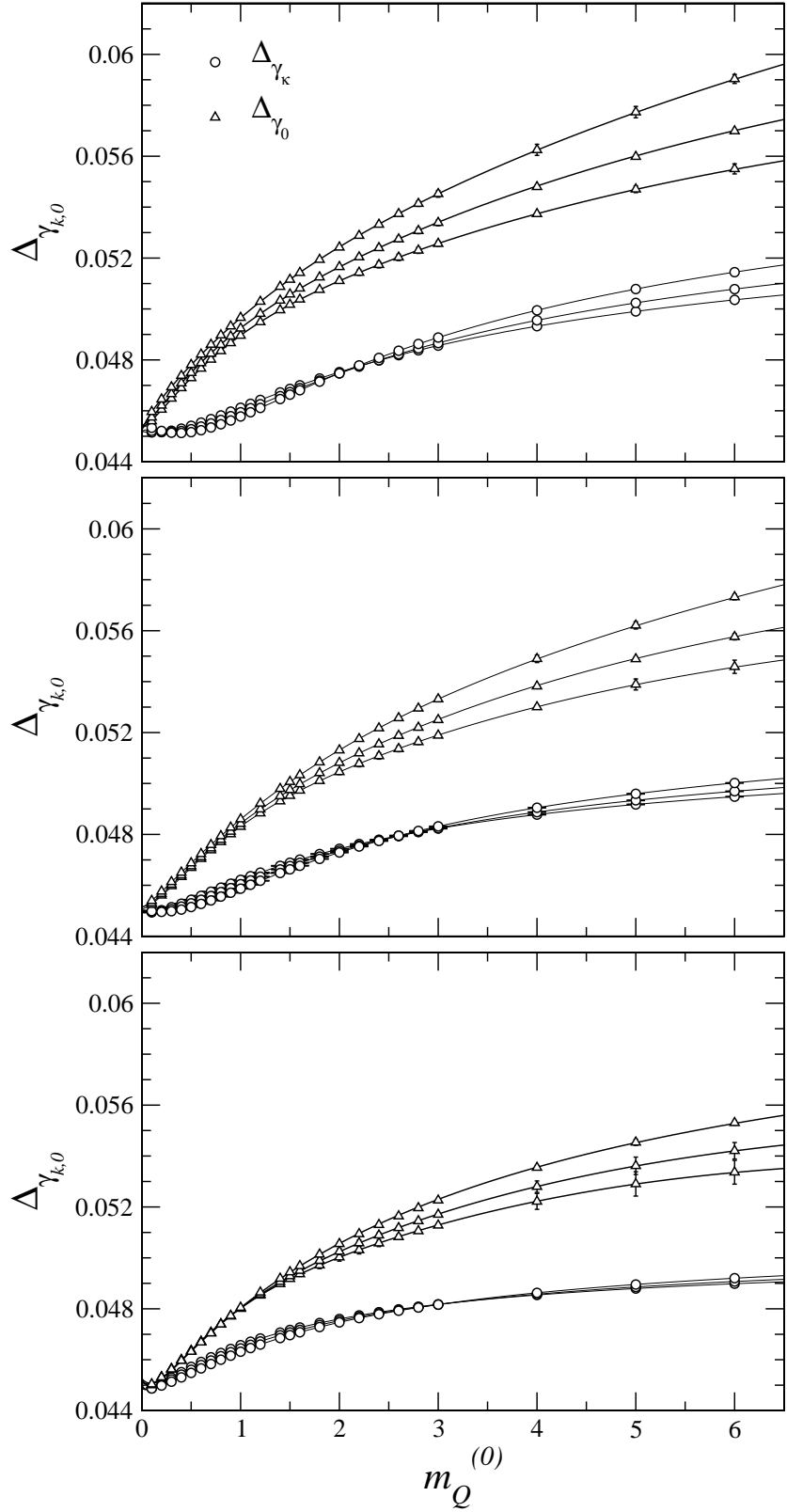

 \begin{center}
  \includegraphics*[width=11cm,clip=true]
  {plaq_del-gk0.eps}\\
  \includegraphics*[width=11cm,clip=true]
  {iwasaki_del-gk0.eps}\\
  \includegraphics*[width=11cm,clip=true]
  {dbw2_del-gk0.eps}
 \end{center}
 \caption{The $m_Q^\lo$ dependence of $\Delta_{\gamma_k}$ (circles) and
 $\Delta_{\gamma_0}$ (triangles) with the plaquette, the Iwasaki and the
 DBW2 gauge actions from top to bottom.}
 \label{fig:v1x}
\end{figure}

\begin{figure}[p]
 \begin{center}
  \includegraphics*[width=11cm,clip=true]
  {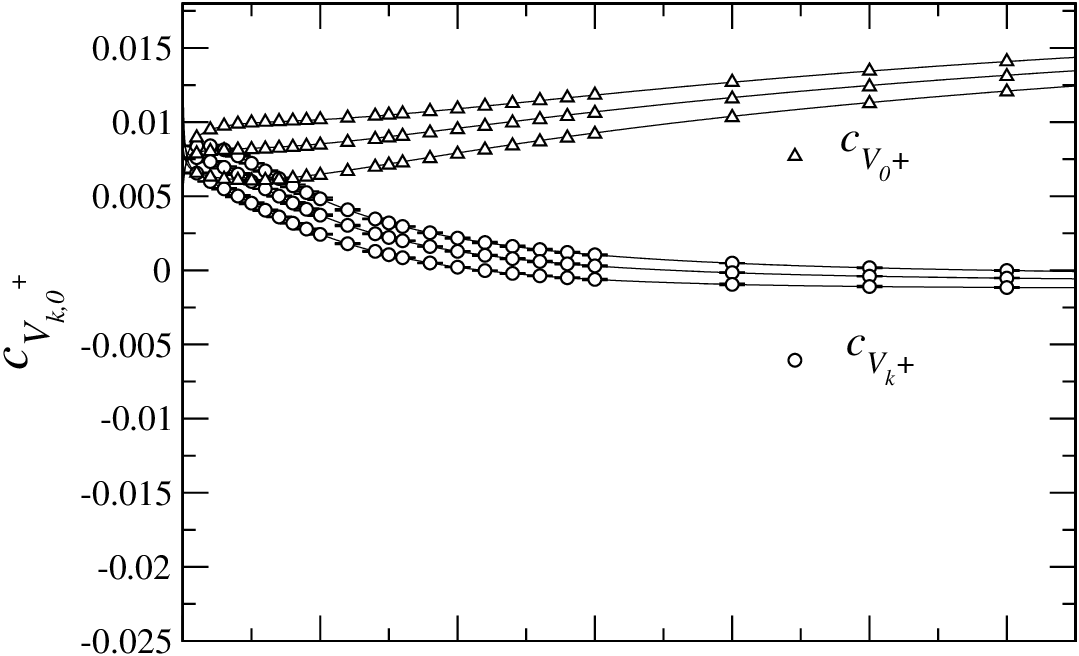}\\
  \includegraphics*[width=11cm,clip=true]
  {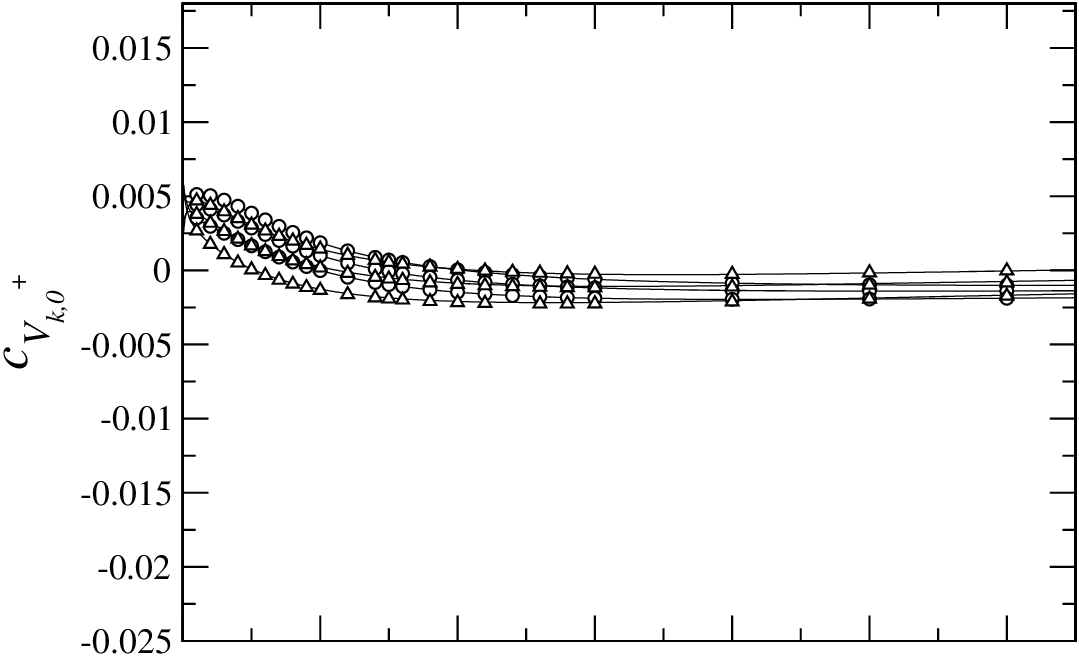}\\
  \includegraphics*[width=11cm,clip=true]
  {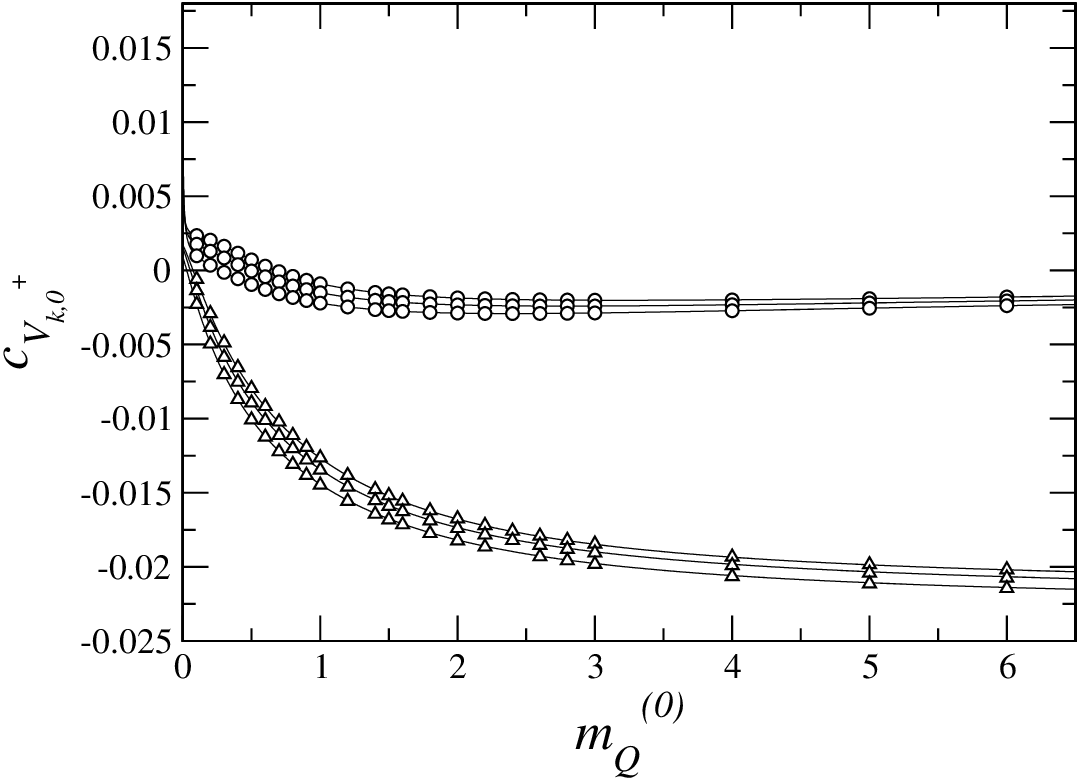}\\
 \end{center}
 \caption{The same figure as Fig.~\ref{fig:v1x} but for $c_{V_k}^+$
 (circles) and $c_{V_0}^+$ (triangles).}
 \label{fig:v1y}
\end{figure}

\begin{figure}[p]
 \begin{center}
  \includegraphics*[width=11cm,clip=true]
  {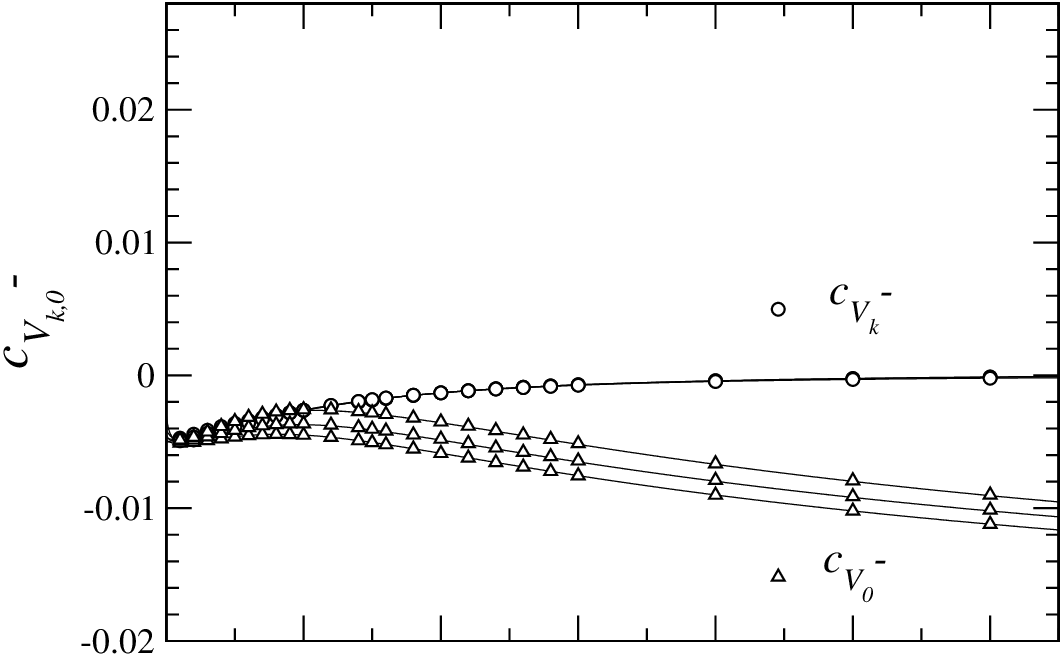}\\
  \includegraphics*[width=11cm,clip=true]
  {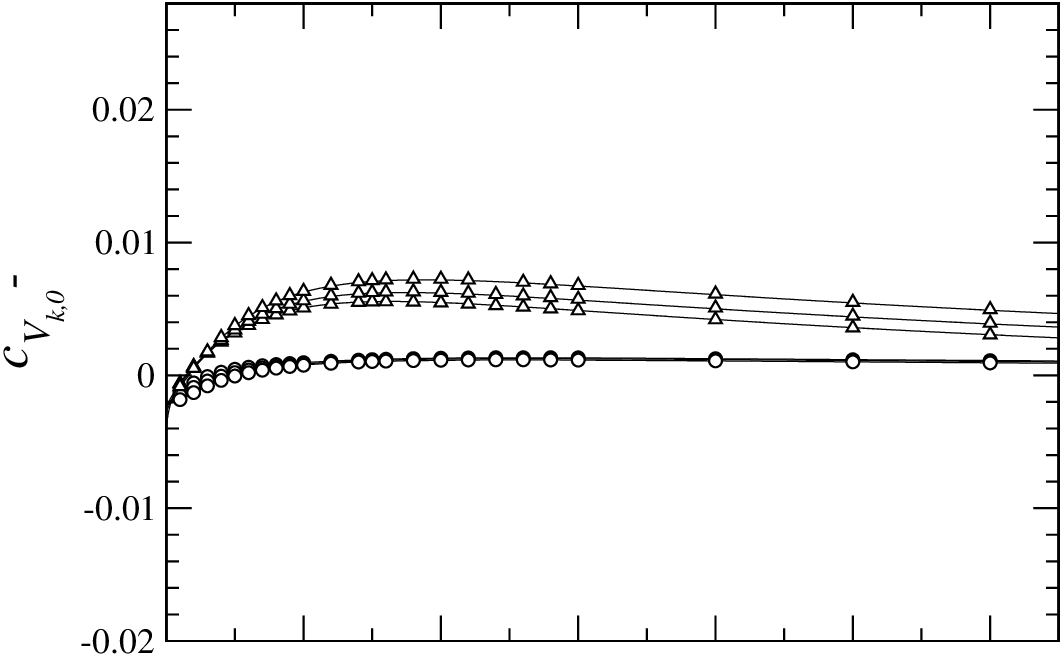}\\
  \includegraphics*[width=11cm,clip=true]
  {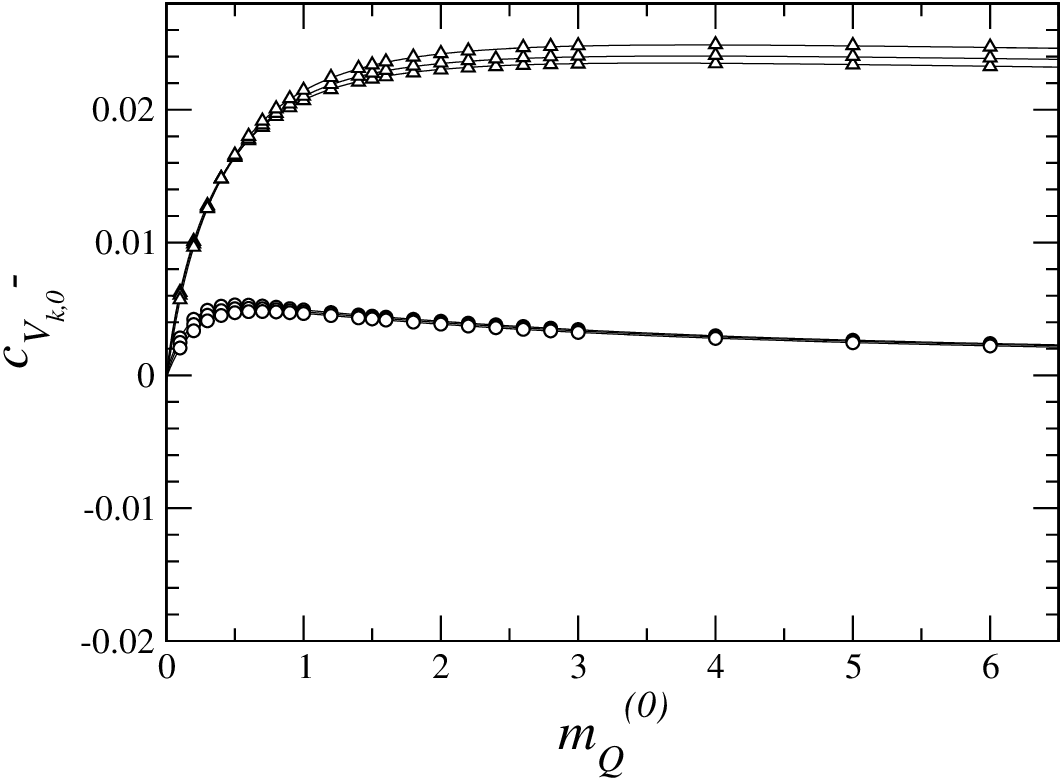}\\
 \end{center}
 \caption{The same figure as Fig.~\ref{fig:v1x} but for $c_{V_k}^-$
 (circles) and $c_{V_0}^-$ (triangles).}
 \label{fig:v1z}
\end{figure}

Figures \ref{fig:v1x}-\ref{fig:v1rs} show the $\mphlo$
dependence of $\Delta_{\gamma_{k,0}}$, $c_{V_{k,0}}^+$, $c_{V_{k,0}}^-$
and $c_{V_k}^{H,L}$, respectively.
In each figure the results with the plaquette, the Iwasaki and DBW2
gauge actions are laid from top to bottom.
We pick up the results with $M_5$=0.5, 1.1 and 1.7 to give some idea
about the $M_5$ dependence, which is turned out to be smaller than the
$m_Q$ dependence in the most cases.
The solid lines in each figure are obtained by fitting the numerical
data to
\begin{eqnarray}
           C
         + \frac{    \sum_{m=1}^{4} N_m\times (\mph^\lo)^m}
                {1 + \sum_{n=1}^{4} D_n\times (\mph^\lo)^n},
 \label{eq:fit_form}
\end{eqnarray}
where $C$, $N_m$ ($m$=1,$\dots$,4) and $D_n$($n$=1,$\dots$,4) are again
parameterized by
\begin{eqnarray}
 C     &=& \sum_{i=0}^{4} c^{(i)} (M_5)^i,\\
 N_m &=& \sum_{i=0}^{4} n_m^{(i)} (M_5)^i,\\
 D_n &=& \sum_{i=0}^{4} d_n^{(i)} (M_5)^i.
  \label{eq:polyfit}
\end{eqnarray}
45 parameters are thus determined by the fit.
Their numerical values are tabulated in Tabs.~\ref{tab:v1x}-\ref{tab:v1s}
for each improvement coefficient with three different gauge actions.
The fit lines reproduce the data points very well over the whole region
of $m_Q^\lo$ and $M_5$ as seen in the figures.

In the limit of $\mphlo=\mpllo$=0, the presence of space-time rotational
symmetry guarantees that $c_{V_k}^{H,L}=0$ independently of the gauge
action and $M_5$.
So we put the constraint of $c^{(i)}$=0 ($i$=0,1,2,3,4) in the fit of
these coefficients.
The space-time rotational symmetry also guarantees that
$\Delta_{\gamma_k}=\Delta_{\gamma_0}$ and
$c_{V_k}^{\pm}=c_{V_0}^{\pm}$ at $\mphlo=\mpllo=0$.
These relations are not used in the fits as constraints, but they are
found to be satisfied within 1 \% in any cases.

We also calculated the renormalization and improvement coefficients for
the axial-vector currents independently, and confirmed that the
relations derived in
eqs.~(\ref{eq:chiral-relation-start})-(\ref{eq:chiral-relation-end})
well hold within the expected numerical accuracy.

 \section{improving perturbative series}
 \label{sec:imp-of-ptseries}

 \subsection{mean field improvement}
 \label{subsec:mean-field-imp}

 Applying the mean field improvement~\cite{Lepage:1992xa} is expected
 to accelerate convergence of perturbative series.
 In this section we examine the mean field improvement using the results
 obtained in the previous section.
 Following Ref. \cite{Aoki:2003dg}, the improvement of the wave function
 renormalization for the relativistic heavy quark is implemented by
 replacing eq.~(\ref{eq:z_akt-1-loop}) with
 \begin{eqnarray}
     Z^{\rm MF}_Q(\mphlo) = 
     u^{\rm MC}\,Z_{Q,\latt}^\lo(\mphlomf)
     \left[ 1 + g^2
                \bigg( - \frac{C_F}{16\pi^2}\ln(a^2\mu^2)
                       + \Delta_Q(m_Q^\lo)
                       + \frac{C_F}{2}T_{\rm MF}
                       + \frac{\partial \ln Z_{Q,\latt}^\lo}
                              {\partial\mphlo}\Delta m_Q
                 \bigg)
     \right],
 \end{eqnarray}
 where
 \begin{eqnarray}
        \mphlomf
  &=&   \mphlo
      + g^2\frac{C_F}{2}T_{\rm MF}
        \frac{   \sinh(\mphlomf)
               + r_t(\cosh (\mphlomf)-1)
               + 3(r_s(\mphlomf)-1)}
                 {\cosh (\mphlomf)+ r_t\sinh (\mphlomf)} \no\\
  &=& \mphlo
      + g^2\frac{C_F}{2}T_{\rm MF}
        \frac{   \sinh(\mphlo)
               + r_t(\cosh (\mphlo)-1)
               + 3(r_s(\mphlo)-1)}
                 {\cosh (\mphlo)+ r_t\sinh (\mphlo)}
      + O(g^4)
                 \no\\
  &=& \mphlo - g^2 \Delta m_Q + O(g^4),
 \end{eqnarray}
 and $u^{\rm MC}$ is the mean plaquette value measured in a Monte Carlo
 simulation whose perturbative expression is
 \begin{eqnarray}
    u^{\rm PT}
  = 1-g^2\frac{C_F}{2} T_{\rm MF},
 \end{eqnarray}
 where $T_{\rm MF}$= 1/8, 0.0525664 and 0.0191580 for the plaquette, the
 Iwasaki and DBW2 action, respectively.
 The improved wave function renormalization for the the domain-wall
 fermions is given by~\cite{Aoki:2002iq}
 \begin{eqnarray}
   \frac{Z^{\rm MF}_q(\tilde{w}_0)}
        {(1-\tilde{w}_0^2)\,Z^{\rm MF}_w(\tilde{w}_0)}
 = u^{\rm MC}
   \frac{\displaystyle
         1 + g^2\left( - \frac{C_F}{16\pi^2}\ln(a^2\mu^2)
                       + \Delta_q(\tilde{M}_5)
                       + \frac{C_F}{2}T_{\rm MF}
                \right)}
        {(1-\tilde{w}_0^2)\,
         \bigg[\ 1 + g^2 \big(\ \Delta_w(\tilde{w}_0)
                          + 2\,C_F\,T_{\rm MF}
                         \big)\
         \bigg]
        },
 \end{eqnarray}
 where $\tilde{w}_0=1-\tilde{M}_5=1-\left( M_5 - 4(1-u)\right)$.

 Now taking into account the above reorganizations of the wave function
 renormalizations, $\mphlo$ and $M_5$, the mean field improved
 renormalization factor is given by
 \begin{eqnarray}
     Z_{V_\mu}^{\msbar-\latt,{\rm MF}}
 &=& u^{\rm MC}\,
     \sqrt{\frac{Z_{Q,\latt}^\lo(\mphlomf)}
          {(1-\tilde{w}_0)\,Z^{\rm MF}_w(\tilde{w}_0)}}
     \bigg[ 1 - g^2{\Delta}^{\rm MF}_{V_\mu} \bigg],
     \label{eq:vmu-tad}\\
     \Delta^{\rm MF}_{V_\mu}
 &=&    \Delta_{V_\mu}(\mphlomf,\tilde{w}_0)
     - \frac{C_F}{2}T_{\rm MF}
     - \frac{1}{2}\frac{\partial\,\ln Z_{Q,\latt}^\lo}
                       {\partial \mphlo}\Delta \mph.
 \end{eqnarray}

 \subsection{quasi non-perturbative improvement}
 \label{subsec:quasi-np-imp}

 Let $Z_{\bar{Q}Q,V_0}^{\msbar-\latt}$ be the overall
 renormalization factor for the lattice heavy-heavy vector current
 $\bar{Q}\gamma_0 Q$.
 Similarly define $Z_{\bar{q}q,V}^{\msbar-\latt}$ for the
 light-light vector current $\bar{q}\gamma_0 q$.
 These factors can be determined non-perturbatively without any special
 devises.
 We can, then, reorganize the perturbative series eq.~(\ref{eq:z_v}) by
 using one or both of these factors as done in Ref.~\cite{Harada:2001fi}.
 For example, if one uses the non-perturbative value of
 $Z_{\bar{q}q,V}^{\msbar-\latt,\rm NP}$
 together with the mean field improvement, eq.~(\ref{eq:z_v}) is
 modified as
 \begin{eqnarray}
     Z_{V_\mu}^{\msbar-\latt,{\rm QNP+MF}}
 &=& \sqrt{u^{\rm MC}}\sqrt{Z^{(0)}_{Q,\latt}(\mphlomf)
           Z_{\bar{q}q,V}^{\msbar-\latt,\rm NP}(w_0)}
     \bigg[ 1 - g^2\Delta^{\rm QNP+MF}_{V_\mu} \bigg],\\
     \Delta^{\rm QNP+MF}_{V_\mu}
 &=&   \Delta_{V_\mu}(\mphlomf,\tilde{w}_0)
     - \frac{1}{2}\Delta_{V}^{\rm LL}(\tilde{w}_0)
     - \frac{C_F}{4}T_{\rm MF}
     - \frac{1}{2}\frac{\partial\,\ln Z_{Q,\latt}^\lo}
                       {\partial \mphlo}\Delta \mph,
     \label{eq:qnp-hmf-l}
 \end{eqnarray}
 where the perturbative expression
 \begin{eqnarray}
 Z_{\bar{q}q,V}^{\msbar-\latt}(w_0)
  = 1 + g^2\,\Delta_{V}^{\rm LL}(w_0),
 \end{eqnarray}
 is used, and
 $\Delta_{V}^{\rm LL}(w_0)$ is given by
 \begin{eqnarray}
      \Delta_{V}^{\rm LL}(w_0)
  &=& -\frac{C_F}{16\pi^2}z_{V/A}({\rm NDR})
   =  -\frac{C_F}{16\pi^2}\left(\, z_{V/A}({\rm DRED})-\frac{1}{2}
                          \right).
 \end{eqnarray}
 The numerical values of $z_{V/A}({\rm DRED})$
 are available in Tables I, III and V of Ref.~\cite{Aoki:2002iq}.
 This way of the improvement has an advantage that the factor
 $Z_w(w_0)$ is automatically included in
 $Z_{\bar{q}q,V}^{\msbar-\latt,\rm NP}(w_0)$.

\begin{figure}[p]
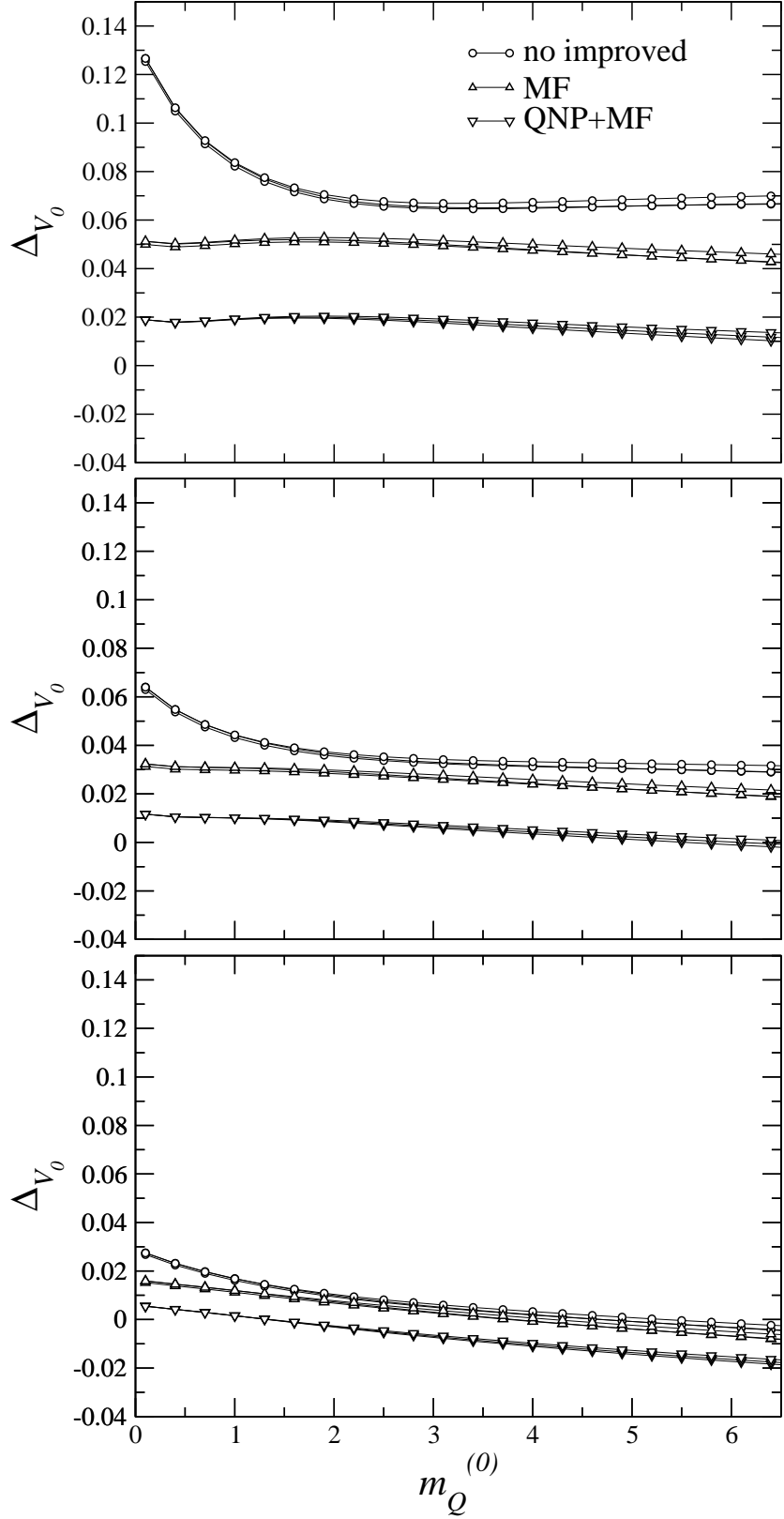

 \begin{center}
  \includegraphics*[width=11cm,clip=true]
  {plaq_del-vk0.eps}\\
  \includegraphics*[width=11cm,clip=true]
  {iwasaki_del-vk0.eps}\\
  \includegraphics*[width=11cm,clip=true]
  {dbw2_del-vk0.eps}
 \end{center}
 \caption{The $m_Q^\lo$ dependence of $\Delta_{V_{0}}$,
 $\Delta_{V_{0}}^{\rm MF}$ and $\Delta_{V_{0}}^{\rm QNP+MF}$ with the
 plaquette, Iwasaki and DBW2 gauge action from top to bottom.
 }
 \label{fig:zvk0}
\end{figure}

 We plot $\Delta_{V_{0}}$, $\Delta_{V_{0}}^{\rm MF}$ and
 $\Delta_{V_{0}}^{\rm QNP+MF}$ in Fig.~\ref{fig:zvk0}, where the results
 with $M_5$=0.5, 1.1 and 1.7 for three gauge actions are shown as
 before.
 From the figure, it is seen that in the non-improved case the size of
 the one-loop coefficient depends on the gauge action while once one
 introduced the quasi non-perturbative and mean field improvements
 together the dependence becomes less transparent.
 It is also interesting to see that the mean field improvement makes the
 $m_Q$ dependence smaller.
 The similar findings are made for the improvements of $\Delta_{V_k}$ as
 well.

 \section{conclusion and discussion}
 \label{sec:conlcusion}

 In this paper, we determined the renormalization coefficients for the
 heavy-light vector and axial-vector currents consisting of the
 relativistic heavy and the domain-wall light quarks.
 The improvement coefficients associated with the dimension four operators 
 are also evaluated, so that
 the leading errors are reduced to $O(\alpha_s^2)$ and $O(a^2\p^2)$.
 The results are presented as an interpolating formula with two
 arguments, the heavy quark mass $m_Q$ and the domain-wall height $M_5$.

 Using light quark actions with the exact chiral symmetry brings
 the useful relations between vector and axial-vector current
 renormalization/improvement coefficients.
 One of the relations
 $Z_{V_\mu}^{\msbar-\latt}=Z_{A_\mu}^{\msbar-\latt}$
 is especially useful for the test of the soft pion relation
 $f^0(q^2_{\rm max})$=$f_B/f_\pi$~\cite{Kitazawa:1993bk}, where $f^0(q^2)$ is
 one of the form factors defined in $B\rightarrow \pi l\nu$ transition
 amplitude.
 In this test, the ratio of
 $Z_{V_\mu}^{\msbar-\latt}/Z_{A_\mu}^{\msbar-\latt}$
 plays an important role, and the unknown two loop coefficient in the
 ratio has prevented the test from being more
 precise.
 So once one determined $c^{\pm,H,L}_{V_\mu}$ at the one loop level,
 the leading uncertainty in the test is reduced to $O(\alpha_s^2a\p)$.
 Furthermore the use of the lattice chiral fermions for the light quark
 is expected to make the chiral extrapolation easier than that of the
 Wilson type light quark.
 With the lattice chiral fermions, thus, one can expect the significant
 reduction of the systematic error in the test.

 \section*{Acknowledgments}
 The main part of the calculations were performed at Riken Super
 Combined Cluster System (RSCC) at Riken.
 N.Y. would like to thank the member of the RBC collaboration for useful
 discussions.
 This work is supported in part by the Grants-in-Aid for
 Scientific Research from the Ministry of Education, 
 Culture, Sports, Science and Technology.
 (Nos. 13135204, 15204015, 15540251, 15740165, 16028201.)

 \appendix
\section{on-shell improvement of the massive domain-wall fermions:
         tree-level analysis} 
\label{sec:on-shell-imp-dwf}

 First let us define the following quantities,
 \begin{eqnarray}
     \slabar{p}
 &=& \gamma_0 \sin(p_0) + \nu_q\sum_i\gamma_i\sin(p_i),\\
     \overline{p}^2
 &=& \sin^2(p_0) + \nu_q^2\sum_i\sin^2(p_i),\\
     W(p)
 &=& 1 - M_5 + 2 \sum_\mu \sin^2\left(\frac{p_\mu}{2}\right),\\
     \cosh(\alpha)
 &=& \frac{1+W^2(p) + \overline{p}^2}{2 |W(p)|},\\
     m_f(p)
 &=& m_f + 2\ R_s\sum_i\sin^2\left(\frac{p_i}{2}\right).\\
    \omega
 &=& 2-M_5.
 \end{eqnarray}
 With the modifications made in eqs.~(\ref{eq:newterm1}) and
 (\ref{eq:newterm2}), the physical quark propagator with momentum $p$
 is given by
 \begin{eqnarray}
     S_q(p)
 &=& \frac{- i\,\slabar{p} + m_f(p)\,(\,1-W(p)\,e^{-\alpha})}
          {  (\,W(p)\,e^\alpha-1\,) + m_f^2(p)\,(\,1-W(p)\,e^{-\alpha})}.
 \end{eqnarray}
 The pole mass is obtained by finding the position of pole in the
 complex $p_0$ plane with $p_i$=0, and given by
 \begin{eqnarray}
   m^\lo_q = \ln\left|\frac{-m_f\,\omega^2
                       +\sqrt{(1+m_f^2)^2+m_f^2\,\omega^2(\omega^2-4)}}
                      {1+m_f^2-2m_f\,\omega}\right|.
\label{eq:polemass}
 \end{eqnarray}
 In the case of $m_f\ll$1, the pole mass reduces to
 \begin{eqnarray}
     m_q^\lo
   = M_5\,(2-M_5)\,m_f + O(m_f^3).
 \end{eqnarray}
 Note that the correction starts from $O(m_f^2)$ relative to the
 leading one. This is consistent with the discussion of chiral
 symmetry made in Ref.~\cite{Aoki:2001ra}.
 Solving this inversely, the bare mass $m_f$ can be written as a
 function of the pole mass as
 \begin{eqnarray}
  m_f = \frac{   \omega  \,(2\,\chq-\omega)
               - \sqrt{  \omega^2\,(2\,\chq-\omega)^2
                       - 4\,(\shq)^2}}
             {2\,\shq}.
  \label{eq:mf-mpole}
 \end{eqnarray}

 Defining the wave function renormalization of the DW quark field
 $Z_q$ by
 \begin{eqnarray}
     q_R = Z_{q,\latt}^{1/2}\ q_{\latt},
 \end{eqnarray}
 its tree level value,
 \begin{eqnarray}
      Z^\lo_{q,\latt}
  &=& \frac{m_f\,(1+m_f^2)\,\chq -2m_f^2\,\omega\,\shq}
           {(1-m_f^2)\,\shq},
 \label{eq:wavefunc}
 \end{eqnarray}
 is obtained by calculating residue of propagator at $p_0=im_q^\lo$.
 In the limit of $m_f\ll$1, a well-known factor
  \begin{eqnarray}
           Z^\lo_{q,\latt}
   \approx \frac{m_f}{m_q^\lo}
   \approx \frac{1}{M_5 (2-M_5)} + O(m_f^2),
  \end{eqnarray}
  is recovered.

 $\nu_q^\lo$ and $R_s^\lo$ are determined by imposing the correct dispersion
 relation on the quark propagator.
 From the denominator,
 it is found that $\nu_q^\lo$ and $R_s^\lo$ must satisfy
 \begin{eqnarray}
 & &   (\nu^\lo_q)^2\,\left(\,1 + m_f^2-2\,m_f\,\shq\,\right)
     + R_s^\lo\,\frac{(\shq)^2}{m_f}(\,1 - m_f^2\,)\no\\
 &=&   \frac{\shq}{m_q^\lo}
       \left[\,(\,1+ m_f^2\,)\,\chq - 2\,m_f\,\omega\,\shq\,
       \right]
     + 2\,m_f\,\left( \omega - \chq \right),
     \label{eq:condition-nu-Rs}
 \end{eqnarray}
 while from the numerator we obtain eq.~(\ref{eq:nuq}).
 Substituting eq.~(\ref{eq:nuq}) to eq.~(\ref{eq:condition-nu-Rs}),
 eq.~(\ref{eq:R_s-tune}) is derived.

 We can further extend the improvement by adding the clover terms to the
 action, which removes the $O((am_q)^n a\p)$ errors thoroughly, though
 this is not necessary in the present work of the main text.
 Although there are several possible ways to include the clover terms,
 we choose to add the following form,
 \begin{eqnarray}
& & - \bar{q}(x)\bigg[
        C_{E,q} \frac{i g}{2}\sum_i    \sigma_{0i}F_{0i}
      + C_{B,q} \frac{i g}{4}\sum_{i,j}\sigma_{ij}F_{ij}
             \bigg] q(x)\no\\
&=& - \bar{\psi}_s(x)\ \bigg[
        C_{E,q} \frac{i g}{2}\sum_i    \sigma_{0i}F_{0i}
      + C_{B,q} \frac{i g}{4}\sum_{i,j}\sigma_{ij}F_{ij}
                       \bigg]
      \left(   P_L\ \delta_{s,N_5}\delta_{t,1}
             + P_R\ \delta_{s,1}\delta_{t,N_5}
      \right)\ \psi_t(x),
   \label{eq:clover}
 \end{eqnarray}
 to the action eq.~(\ref{eq:action_dw}).
 To tune the new parameters $C_{E,q}$ and $C_{B,q}$, it is enough to
 consider a quark form factor scattering off a single
 gluon~\cite{Aoki:2003dg}.
 The tree level form factor $F_\mu^a(p_1,p_2)$ is, then, given by
 \begin{eqnarray}
      F^a_0(p_1,p_2)
  =&&\!\!\!
      -i g\ (Z_{q,\latt}^{(0)})^{-1} \times\bigg[\
      \bigg\{\ \frac{m_f\,\cosh(m_q^\lo)\,
                     \bigg(1+m_f^2-2m_f\,\sinh(m_q^\lo)\bigg)}
                    {\sinh(m_q^\lo)\,(1-m_f^2)}\,
             +\ \frac{C_{E,q}^\lo}{\nu_q^\lo}\,\sinh(m_q^\lo)
      \bigg\}
    \no\\
   &&\hspace{19ex}
      \bar{u}_q(p_1)\gamma_0 T^a u_q(p_2)\no\\
   &&\hspace{16ex}
    -\ \bigg\{\ \frac{2m_f^2\left(\omega-\cosh(m_q^\lo)\right)}
                     {1-m_f^2}
            + \ \frac{C_{E,q}^\lo}{\nu_q^\lo}\
                \sinh(m_q^\lo)  
       \bigg\}\,
            \bar{u}_q(p_1) T^a u_q(p_2)
           \bigg]\
       \no\\
   &&\hspace{16ex}
           + O(a^2\p_{1,2}^2),
\label{eq:ce-tune}
 \end{eqnarray}
 for $\mu=0$, and
 \begin{eqnarray}
      F^a_i(p_1,p_2)
  =&&\!\!\!
      -i g\ (Z_{q,\latt}^{(0)})^{-1} \times\bigg[\
      \bigg\{\  \frac{\nu_q^\lo\,m_f\,\bigg(1+m_f^2-2m_f\sinh(m_q^\lo)\bigg)}
                     {\sinh(m_q^\lo)\,(1-m_f^2)}
             +\ \frac{C_{B,q}^\lo}{\nu_q^\lo}\ \sinh(m_q^\lo)
      \bigg\}\no\\
   &&\hspace{21ex}
      \bar{u}_q(p_1)\gamma_i T^a u_q(p_2)\no\\
   &&\hspace{18ex}
      +\ i({p_1}_i + {p_2}_i)
         \bigg\{
          \frac{m_f^2\left(\omega-\cosh(m_q^\lo)\right)}
               {\sinh(m_q^\lo)\,(1-m_f^2)}
        - \frac{R_s^\lo}{2}
        + \frac{C_{B,q}^\lo}{2}
         \bigg\}\no\\
   &&\hspace{21ex} \times
           \bar{u}_q(p_1) T^a u_q(p_2)
           \bigg],
  \label{eq:cb-tune}
 \end{eqnarray}
 for $\mu=i$.
 Now using eq.~(\ref{eq:nuq}), (\ref{eq:R_s-tune}) and
 (\ref{eq:wavefunc}), it is turned out that setting 
 \begin{eqnarray}
     C_{E,q}^\lo
 &=& -\,\frac{2 m_f^2 \left( \omega - \cosh(m_q^\lo) \right)}
             {m_q^\lo\,( 1 - m_f^2)},\\
     C_{B,q}^\lo
 &=& \frac{m_f\,(1+m_f^2)
                \left(  \cosh(m_q^\lo)
                      - \frac{\displaystyle\sinh(m_q^\lo)}
                             {\displaystyle m_q^\lo}
                \right)          
          - 2\ m_f^2\sinh(m_q^\lo)\
            \left(\omega -\frac{\displaystyle\sinh(m_q^\lo)}
                               {\displaystyle m_q^\lo}
            \right)}
            {m_q^\lo\,\sinh(m_q^\lo)\,(1-m_f^2)},
  \label{eq:cbq}
 \end{eqnarray}
 reproduce the correct form factor,
 \begin{eqnarray}
    F^a_\mu(p_1,p_2)
  = - i g\ \bar{u}_q(p_1)\gamma_\mu T^a u_q(p_2) + O(a^2\p_{1,2}^2).
 \end{eqnarray}
 Note that
 $C_{E,q}^\lo$ and $C_{B,q}^\lo$ vanish in the massless limit,
 which means that the action does not have $O(a)$ errors in this limit.
 This is consistent with the presence of chiral symmetry in the
 domain-wall fermion.

 \bibliography{basename of .bib file}

 \newcommand{\cc}[1]{\multicolumn{5}{c}{#1}}
\renewcommand{\arraystretch}{1}

\clearpage
\begin{table}[htb]
 \begin{tabular}{c|rrrrr}
   $i$&0&1&2&3&4\\
   \hline
   % Fit results for v1x_1a1x with plaquette
& \cc{plaquette}\\ 
$c^{(i)}$ &   4.50349e$-$02 &   6.08797e$-$04 &  $-$6.18501e$-$04 &   2.78999e$-$05 &   9.79888e$-$05 \\
$n_{1}^{(i)}$ &   2.19429e$-$03 &  $-$1.16680e$-$02 &  $-$4.02314e$-$03 &   3.75624e$-$02 &  $-$1.62773e$-$02 \\
$n_{2}^{(i)}$ &   3.20289e$-$03 &   3.14583e$-$02 &  $-$1.83939e$-$01 &   1.22658e$-$01 &  $-$3.12492e$-$02 \\
$n_{3}^{(i)}$ &   1.06077e$-$02 &   8.31176e$-$02 &   2.83741e$-$02 &  $-$1.18936e$-$02 &  $-$7.77838e$-$03 \\
$n_{4}^{(i)}$ &   3.89916e$-$04 &   6.63305e$-$02 &  $-$2.93231e$-$02 &   2.14512e$-$02 &  $-$4.35142e$-$03 \\
$d_{1}^{(i)}$ &   4.88310e+00 &   4.68459e+01 &  $-$1.26499e+01 &   8.52440e+00 &  $-$3.88139e+00 \\
$d_{2}^{(i)}$ &   4.33555e+00 &   4.37762e+01 &   9.61199e+00 &  $-$1.43073e+01 &   2.71429e+00 \\
$d_{3}^{(i)}$ &   1.53503e+00 &   3.98540e+01 &  $-$6.97580e+00 &   3.70893e$-$01 &  $-$2.30419e$-$01 \\
$d_{4}^{(i)}$ &   4.13266e$-$02 &   7.86025e+00 &  $-$4.21872e+00 &   2.14446e+00 &  $-$4.47614e$-$01 \\
\hline
% Fit results for v1x_1a1x with Iwasaki
& \cc{Iwasaki}\\ 
$c^{(i)}$ &   4.48266e$-$02 &   3.98377e$-$04 &  $-$4.08681e$-$04 &   8.38875e$-$05 &   3.25504e$-$05 \\
$n_{1}^{(i)}$ &   2.01108e$-$03 &  $-$1.03944e$-$02 &   6.00284e$-$03 &   8.16515e$-$03 &  $-$4.66134e$-$03 \\
$n_{2}^{(i)}$ &   5.10189e$-$03 &   4.09113e$-$02 &  $-$1.10778e$-$01 &   6.81677e$-$02 &  $-$1.67361e$-$02 \\
$n_{3}^{(i)}$ &   5.50012e$-$03 &   6.34936e$-$02 &   6.54794e$-$02 &  $-$4.65406e$-$02 &   3.99199e$-$03 \\
$n_{4}^{(i)}$ &  $-$3.75123e$-$04 &   2.81289e$-$02 &  $-$1.68894e$-$02 &   1.30670e$-$02 &  $-$2.55968e$-$03 \\
$d_{1}^{(i)}$ &   2.98626e+00 &   2.57075e+01 &  $-$3.24294e+00 &   5.57942e+00 &  $-$3.34665e+00 \\
$d_{2}^{(i)}$ &   2.70854e+00 &   2.82522e+01 &   1.82688e+01 &  $-$1.70378e+01 &   2.77775e+00 \\
$d_{3}^{(i)}$ &   7.22072e$-$01 &   2.16565e+01 &   4.64775e+00 &  $-$5.30941e+00 &   8.56825e$-$01 \\
$d_{4}^{(i)}$ &  $-$5.90865e$-$02 &   4.25941e+00 &  $-$2.83516e+00 &   1.86689e+00 &  $-$3.96036e$-$01 \\
\hline
% Fit results for v1x_1a1x with DBW2
& \cc{DBW2}\\ 
$c^{(i)}$ &   4.47375e$-$02 &   5.75118e$-$04 &  $-$3.46013e$-$04 &   8.99787e$-$05 &   9.51994e$-$06 \\
$n_{1}^{(i)}$ &   1.91634e$-$03 &  $-$2.99489e$-$02 &   2.04894e$-$02 &  $-$3.25490e$-$02 &   1.00940e$-$02 \\
$n_{2}^{(i)}$ &   4.13973e$-$02 &   9.03824e$-$02 &  $-$2.27829e$-$01 &   2.26370e$-$01 &  $-$8.34536e$-$02 \\
$n_{3}^{(i)}$ &   6.61204e$-$02 &   2.46242e$-$01 &   2.16915e$-$01 &  $-$8.46906e$-$02 &  $-$9.70525e$-$03 \\
$n_{4}^{(i)}$ &  $-$3.73368e$-$03 &   6.80095e$-$02 &  $-$6.85793e$-$02 &   5.41115e$-$02 &  $-$1.24966e$-$02 \\
$d_{1}^{(i)}$ &   1.32735e+01 &   5.46203e+01 &  $-$2.94128e+01 &   7.78518e+01 &  $-$3.16231e+01 \\
$d_{2}^{(i)}$ &   2.63415e+01 &   8.32744e+01 &   6.80241e+01 &  $-$5.89169e+00 &  $-$1.19411e+01 \\
$d_{3}^{(i)}$ &   1.15204e+01 &   6.95572e+01 &   2.59625e+01 &  $-$5.71156e+00 &  $-$3.67863e+00 \\
$d_{4}^{(i)}$ &  $-$7.17896e$-$01 &   1.26939e+01 &  $-$1.26694e+01 &   9.67410e+00 &  $-$2.26554e+00 \\
\hline

   \hline
 \end{tabular}
  \caption{Numerical results of the fit parameters for
  $\Delta_{\gamma_k}$.}
  \label{tab:v1x}
\end{table}
\begin{table}[htb]
 \begin{tabular}{c|rrrrr}
   $i$&0&1&2&3&4\\
   \hline
   % Fit results for v4x_1a4x with plaquette
& \cc{plaquette}\\ 
$c^{(i)}$ &   4.52373e$-$02 &   7.95780e$-$06 &   5.08306e$-$04 &  $-$4.22552e$-$04 &   6.99005e$-$05 \\
$n_{1}^{(i)}$ &   2.84911e$-$03 &   7.82834e$-$04 &  $-$3.98343e$-$03 &   7.84253e$-$04 &   1.64635e$-$03 \\
$n_{2}^{(i)}$ &   1.04294e$-$02 &  $-$3.50230e$-$03 &   2.54349e$-$02 &  $-$8.17546e$-$03 &   2.13300e$-$03 \\
$n_{3}^{(i)}$ &  $-$4.07134e$-$03 &   1.79561e$-$02 &  $-$2.42208e$-$02 &   1.70135e$-$02 &  $-$4.40267e$-$03 \\
$n_{4}^{(i)}$ &   6.34372e$-$04 &  $-$4.76941e$-$03 &   1.51021e$-$02 &  $-$6.08709e$-$03 &  $-$6.67261e$-$05 \\
$d_{1}^{(i)}$ &   1.06023e+00 &   1.31409e+00 &   1.05189e+00 &   7.46079e$-$01 &   2.09868e$-$01 \\
$d_{2}^{(i)}$ &   1.06865e+00 &   8.17638e$-$01 &   7.32570e$-$01 &   4.22780e$-$01 &  $-$5.17816e$-$01 \\
$d_{3}^{(i)}$ &  $-$4.28339e$-$01 &   9.00811e$-$01 &   8.22650e$-$01 &   3.40122e$-$01 &  $-$4.14352e$-$01 \\
$d_{4}^{(i)}$ &   6.74709e$-$02 &  $-$4.02057e$-$01 &   1.18565e+00 &  $-$7.43252e$-$01 &   1.20746e$-$01 \\
\hline
% Fit results for v4x_1a4x with Iwasaki
& \cc{Iwasaki}\\ 
$c^{(i)}$ &   4.49640e$-$02 &   1.43870e$-$04 &   1.38586e$-$04 &  $-$8.90746e$-$05 &  $-$4.85512e$-$06 \\
$n_{1}^{(i)}$ &   2.55443e$-$03 &  $-$1.78947e$-$03 &  $-$1.16834e$-$03 &   7.84991e$-$06 &   6.30484e$-$04 \\
$n_{2}^{(i)}$ &   6.56198e$-$03 &   4.01176e$-$03 &   9.98744e$-$03 &  $-$3.19040e$-$04 &  $-$9.73106e$-$04 \\
$n_{3}^{(i)}$ &  $-$2.32566e$-$03 &   8.53741e$-$03 &  $-$9.25619e$-$03 &   6.13565e$-$03 &  $-$2.25368e$-$03 \\
$n_{4}^{(i)}$ &   2.48873e$-$04 &  $-$1.39051e$-$03 &   5.97769e$-$03 &  $-$2.99920e$-$03 &   2.50580e$-$04 \\
$d_{1}^{(i)}$ &   6.08980e$-$01 &   1.33407e+00 &   9.80322e$-$01 &   5.33446e$-$01 &  $-$1.03361e$-$01 \\
$d_{2}^{(i)}$ &   8.28020e$-$01 &   8.06319e$-$01 &   6.73518e$-$01 &   3.08031e$-$01 &  $-$6.20501e$-$01 \\
$d_{3}^{(i)}$ &  $-$3.09041e$-$01 &   6.39750e$-$01 &   2.78380e$-$01 &  $-$1.60949e$-$01 &  $-$2.75471e$-$02 \\
$d_{4}^{(i)}$ &   3.30870e$-$02 &  $-$1.23565e$-$01 &   4.61016e$-$01 &  $-$2.91189e$-$01 &   4.43351e$-$02 \\
\hline
% Fit results for v4x_1a4x with DBW2
& \cc{DBW2}\\ 
$c^{(i)}$ &   4.49949e$-$02 &  $-$5.33912e$-$05 &   9.92059e$-$04 &  $-$9.12758e$-$04 &   2.31602e$-$04 \\
$n_{1}^{(i)}$ &   1.25755e$-$05 &  $-$2.32660e$-$03 &  $-$9.17352e$-$03 &   1.00097e$-$02 &  $-$3.13132e$-$03 \\
$n_{2}^{(i)}$ &   1.17640e$-$02 &   7.59642e$-$03 &   8.23216e$-$03 &  $-$9.46580e$-$03 &   4.57219e$-$03 \\
$n_{3}^{(i)}$ &  $-$2.35282e$-$03 &   4.99363e$-$03 &  $-$3.60508e$-$03 &   9.55421e$-$03 &  $-$5.03662e$-$03 \\
$n_{4}^{(i)}$ &   7.79286e$-$05 &   5.07874e$-$04 &   7.65744e$-$04 &  $-$1.54716e$-$03 &   5.60927e$-$04 \\
$d_{1}^{(i)}$ &   8.38142e$-$01 &   9.34378e$-$01 &   6.79401e$-$01 &   4.38216e$-$01 &   1.82617e$-$01 \\
$d_{2}^{(i)}$ &   1.79736e+00 &   1.35771e+00 &   9.74925e$-$01 &   3.91138e$-$01 &  $-$7.30343e$-$01 \\
$d_{3}^{(i)}$ &  $-$4.31666e$-$01 &   3.18664e$-$01 &   1.50605e$-$01 &   1.16987e$-$01 &  $-$1.26641e$-$01 \\
$d_{4}^{(i)}$ &   2.39747e$-$02 &   8.19427e$-$02 &   6.10548e$-$04 &  $-$8.72171e$-$02 &   3.34782e$-$02 \\
\hline

   \hline
 \end{tabular}
 \caption{Numerical results of the fit parameters for
          $\Delta_{\gamma_0}$.}
 \label{tab:v4x}
\end{table}
\begin{table}[htb]
 \begin{tabular}{c|rrrrr}
   $i$&0&1&2&3&4\\
   \hline
   % Fit results for v1y_-1a1z with plaquette
& \cc{plaquette}\\ 
$c^{(i)}$ &   9.84230e$-$03 &  $-$3.51914e$-$03 &   6.52579e$-$03 &  $-$2.36438e$-$03 &   3.70171e$-$04 \\
$n_{1}^{(i)}$ &   2.78548e$-$02 &  $-$4.71051e$-$01 &   7.97658e$-$01 &  $-$1.58830e+00 &   5.93228e$-$01 \\
$n_{2}^{(i)}$ &   5.84173e$-$01 &   1.40053e$-$01 &   7.76400e$-$01 &  $-$2.64456e+00 &   9.17475e$-$01 \\
$n_{3}^{(i)}$ &  $-$1.26248e+00 &  $-$1.11733e+00 &  $-$3.07980e+00 &   7.89903e$-$01 &   6.06679e$-$01 \\
$n_{4}^{(i)}$ &   1.52550e$-$01 &  $-$1.26536e+00 &   1.14001e+00 &  $-$9.70884e$-$01 &   2.54240e$-$01 \\
$d_{1}^{(i)}$ &   2.24515e+01 &   1.27087e+02 &   1.13085e+02 &  $-$6.90381e+01 &  $-$6.79194e+00 \\
$d_{2}^{(i)}$ &   5.48959e+01 &   1.19734e+02 &   2.07704e+02 &  $-$1.60331e+02 &   2.14915e+01 \\
$d_{3}^{(i)}$ &   1.05293e+02 &   1.66311e+02 &   2.45242e+02 &  $-$2.24223e+02 &   2.13562e+01 \\
$d_{4}^{(i)}$ &  $-$1.42301e+01 &   1.24790e+02 &  $-$9.07259e+01 &   5.73037e+01 &  $-$1.52792e+01 \\
\hline
% Fit results for v1y_-1a1z with Iwasaki
& \cc{Iwasaki}\\ 
$c^{(i)}$ &   7.10943e$-$03 &  $-$8.49976e$-$03 &   1.02419e$-$02 &  $-$5.11717e$-$03 &   1.09172e$-$03 \\
$n_{1}^{(i)}$ &   9.18767e$-$03 &  $-$7.17314e$-$02 &   2.20257e$-$01 &  $-$2.18720e$-$01 &   5.11634e$-$02 \\
$n_{2}^{(i)}$ &   4.32270e$-$02 &   7.01214e$-$02 &  $-$1.33866e$-$01 &   5.22629e$-$02 &  $-$3.09862e$-$02 \\
$n_{3}^{(i)}$ &  $-$1.61351e$-$01 &  $-$1.04229e$-$01 &  $-$8.88877e$-$03 &  $-$2.63990e$-$01 &   1.15332e$-$01 \\
$n_{4}^{(i)}$ &   5.91548e$-$03 &  $-$3.94141e$-$02 &   5.92655e$-$02 &  $-$9.09445e$-$02 &   2.17575e$-$02 \\
$d_{1}^{(i)}$ &   3.12332e+00 &   1.55819e+01 &   1.37894e+01 &   1.11819e+01 &  $-$1.02538e+01 \\
$d_{2}^{(i)}$ &   8.14387e+00 &   4.36052e+00 &   1.52528e+01 &   7.73470e+00 &  $-$3.22351e+00 \\
$d_{3}^{(i)}$ &   1.83424e+01 &   2.00024e+01 &   1.54535e+01 &   9.90810e+00 &  $-$1.02602e+01 \\
$d_{4}^{(i)}$ &  $-$5.89261e$-$01 &   5.98743e+00 &  $-$7.24667e+00 &   1.55318e+01 &  $-$5.12679e+00 \\
\hline
% Fit results for v1y_-1a1z with DBW2
& \cc{DBW2}\\ 
$c^{(i)}$ &   3.24314e$-$03 &   2.52054e$-$03 &   4.44431e$-$03 &  $-$3.08451e$-$03 &   6.31374e$-$04 \\
$n_{1}^{(i)}$ &   1.05779e$-$01 &  $-$1.01997e+00 &   1.35098e+00 &  $-$2.36283e+00 &   9.06106e$-$01 \\
$n_{2}^{(i)}$ &   1.34493e$-$01 &   7.58544e$-$01 &  $-$1.69656e+00 &   6.16042e$-$01 &  $-$1.90720e$-$01 \\
$n_{3}^{(i)}$ &  $-$7.86597e$-$01 &  $-$2.38384e+00 &   4.72923e$-$03 &  $-$2.82519e+00 &   1.72544e+00 \\
$n_{4}^{(i)}$ &   6.41759e$-$02 &  $-$2.48777e$-$01 &  $-$6.18721e$-$02 &   1.10141e$-$01 &  $-$5.58659e$-$02 \\
$d_{1}^{(i)}$ &   4.67326e+01 &   6.66294e+01 &   9.21788e+00 &   1.05657e+02 &  $-$6.26751e+01 \\
$d_{2}^{(i)}$ &  $-$1.58592e+01 &   7.51684e+01 &   3.20230e+01 &  $-$4.75731e+01 &   2.29401e+01 \\
$d_{3}^{(i)}$ &   1.66161e+02 &   1.31914e+02 &   1.13563e+02 &   4.16193e+01 &  $-$7.63390e+01 \\
$d_{4}^{(i)}$ &  $-$1.23813e+01 &   8.96851e+01 &  $-$8.01392e+01 &   3.67380e+01 &  $-$5.42257e+00 \\
\hline

   \hline
 \end{tabular}
 \caption{Numerical results of the fit parameters for $c_{V_k}^+$.}
 \label{tab:v1y}
\end{table}
\begin{table}[htb]
 \begin{tabular}{c|rrrrr}
   $i$&0&1&2&3&4\\
   \hline
   % Fit results for v4y_-1a4z with plaquette
& \cc{plaquette}\\ 
$c^{(i)}$ &   4.72138e$-$03 &   1.18712e$-$02 &  $-$3.70767e$-$02 &   3.64417e$-$02 &  $-$1.05757e$-$02 \\
$n_{1}^{(i)}$ &   2.88129e$-$01 &  $-$1.02535e+00 &   2.46673e+00 &  $-$2.18206e+00 &   5.91146e$-$01 \\
$n_{2}^{(i)}$ &   1.94450e$-$01 &   6.18542e$-$02 &   5.07036e$-$02 &  $-$2.53440e$-$01 &   6.96176e$-$02 \\
$n_{3}^{(i)}$ &   1.12189e$-$01 &  $-$7.65970e$-$01 &   2.08763e+00 &  $-$1.87971e+00 &   5.16320e$-$01 \\
$n_{4}^{(i)}$ &   6.00100e$-$02 &   5.83904e$-$02 &   5.29609e$-$03 &  $-$7.12338e$-$02 &   3.05537e$-$02 \\
$d_{1}^{(i)}$ &   2.63902e+01 &   1.69601e+01 &   5.43255e+00 &  $-$1.02071e+01 &   4.19682e+00 \\
$d_{2}^{(i)}$ &   1.94234e+01 &   2.93848e+01 &   1.60899e+00 &  $-$2.29237e+01 &   1.07780e+01 \\
$d_{3}^{(i)}$ &   1.26822e+01 &   3.48188e+01 &   5.02924e+00 &  $-$3.22997e+01 &   1.47403e+01 \\
$d_{4}^{(i)}$ &   2.76062e+00 &   1.15866e+01 &  $-$1.75853e+01 &   1.09995e+01 &  $-$2.31401e+00 \\
\hline
% Fit results for v4y_-1a4z with Iwasaki
& \cc{Iwasaki}\\ 
$c^{(i)}$ &   1.53152e$-$03 &   1.03025e$-$02 &  $-$3.42414e$-$02 &   3.46677e$-$02 &  $-$1.00368e$-$02 \\
$n_{1}^{(i)}$ &   3.09889e$-$01 &  $-$1.04777e+00 &   2.55683e+00 &  $-$2.28807e+00 &   6.09905e$-$01 \\
$n_{2}^{(i)}$ &  $-$6.99951e$-$02 &  $-$4.40763e$-$01 &   1.04528e+00 &  $-$1.07999e+00 &   2.66273e$-$01 \\
$n_{3}^{(i)}$ &   4.81714e$-$02 &  $-$6.08612e$-$01 &   1.06819e+00 &  $-$7.59663e$-$01 &   1.48689e$-$01 \\
$n_{4}^{(i)}$ &   7.12031e$-$03 &  $-$4.51084e$-$02 &   1.18638e$-$01 &  $-$1.14891e$-$01 &   3.34277e$-$02 \\
$d_{1}^{(i)}$ &   3.28468e+01 &   2.10820e+01 &  $-$1.72327e+00 &  $-$4.85904e+00 &   2.96723e+00 \\
$d_{2}^{(i)}$ &   2.13491e+01 &   3.82549e+01 &   1.27385e+01 &  $-$3.26290e+01 &   1.42484e+01 \\
$d_{3}^{(i)}$ &   7.91264e+00 &   3.20460e+01 &   1.79611e+00 &  $-$3.17287e+01 &   1.56233e+01 \\
$d_{4}^{(i)}$ &   1.42775e+00 &   5.13906e+00 &  $-$9.29549e+00 &   7.44461e+00 &  $-$2.04473e+00 \\
\hline
% Fit results for v4y_-1a4z with DBW2
& \cc{DBW2}\\ 
$c^{(i)}$ &   3.59477e$-$03 &  $-$2.50302e$-$03 &  $-$7.48782e$-$03 &   1.20853e$-$02 &  $-$4.33286e$-$03 \\
$n_{1}^{(i)}$ &  $-$4.65181e$-$03 &  $-$1.36855e$-$01 &   3.58241e$-$01 &  $-$3.41029e$-$01 &   1.01482e$-$01 \\
$n_{2}^{(i)}$ &  $-$1.55086e$-$01 &  $-$1.09153e$-$01 &   2.47996e$-$02 &  $-$3.56761e$-$02 &  $-$1.71446e$-$02 \\
$n_{3}^{(i)}$ &   6.09454e$-$03 &  $-$1.03449e$-$01 &   1.73653e$-$01 &  $-$1.56521e$-$01 &   5.34803e$-$02 \\
$n_{4}^{(i)}$ &  $-$6.32341e$-$03 &  $-$1.77005e$-$01 &   3.75656e$-$02 &   1.07267e$-$01 &  $-$5.56119e$-$02 \\
$d_{1}^{(i)}$ &   5.18966e+00 &   2.00129e+00 &   3.24520e$-$01 &   1.37221e+00 &  $-$1.01591e$-$01 \\
$d_{2}^{(i)}$ &   6.12021e+00 &   7.41829e+00 &   1.22181e+00 &  $-$4.71897e+00 &   3.05091e+00 \\
$d_{3}^{(i)}$ &   1.89231e$-$01 &   4.22273e+00 &  $-$6.55329e$-$02 &  $-$2.04306e+00 &   5.66055e$-$01 \\
$d_{4}^{(i)}$ &   3.67961e$-$01 &   6.56838e+00 &   9.83657e$-$01 &  $-$7.05932e+00 &   3.11498e+00 \\
\hline

   \hline
 \end{tabular}
 \caption{Numerical results of the fit parameters for $c_{V_0}^+$.}
 \label{tab:v4y}
\end{table}
\begin{table}[htb]
 \begin{tabular}{c|rrrrr}
   $i$&0&1&2&3&4\\
   \hline
   % Fit results for v1z_-1a1y with plaquette
& \cc{plaquette}\\ 
$c^{(i)}$ &  $-$3.77901e$-$03 &  $-$5.20628e$-$03 &   6.92389e$-$03 &  $-$3.33440e$-$03 &   5.83119e$-$04 \\
$n_{1}^{(i)}$ &  $-$5.34436e$-$05 &   1.86231e$-$02 &  $-$3.55829e$-$02 &   2.01514e$-$02 &  $-$5.51715e$-$03 \\
$n_{2}^{(i)}$ &   4.10664e$-$03 &  $-$1.02295e$-$02 &   6.70875e$-$02 &  $-$5.09115e$-$02 &   1.34993e$-$02 \\
$n_{3}^{(i)}$ &  $-$7.67466e$-$03 &   1.17796e$-$02 &  $-$3.73396e$-$02 &   2.86610e$-$02 &  $-$5.86918e$-$03 \\
$n_{4}^{(i)}$ &   7.17646e$-$03 &   4.63334e$-$03 &   3.45373e$-$03 &  $-$2.78005e$-$03 &  $-$6.70178e$-$04 \\
$d_{1}^{(i)}$ &   6.73021e$-$01 &   1.34845e+00 &   1.32647e+00 &   9.83849e$-$01 &  $-$6.91234e$-$01 \\
$d_{2}^{(i)}$ &   8.11070e$-$01 &   1.79291e+00 &   1.05732e+00 &   4.24515e$-$01 &   1.04878e$-$01 \\
$d_{3}^{(i)}$ &  $-$2.10312e$-$01 &  $-$3.58696e$-$01 &  $-$2.70077e$-$01 &  $-$8.84022e$-$02 &   1.93194e$-$01 \\
$d_{4}^{(i)}$ &   1.54957e+00 &   2.32307e$-$01 &   5.35889e$-$01 &   5.32861e$-$01 &  $-$5.67951e$-$01 \\
\hline
% Fit results for v1z_-1a1y with Iwasaki
& \cc{Iwasaki}\\ 
$c^{(i)}$ &  $-$8.92573e$-$04 &  $-$4.98763e$-$03 &   5.83872e$-$03 &  $-$2.80997e$-$03 &   4.99720e$-$04 \\
$n_{1}^{(i)}$ &   8.18233e$-$03 &   1.46155e$-$02 &  $-$2.51372e$-$02 &   1.74577e$-$02 &  $-$5.53396e$-$03 \\
$n_{2}^{(i)}$ &  $-$1.08859e$-$02 &  $-$4.35439e$-$03 &   4.12491e$-$02 &  $-$3.05207e$-$02 &   1.01006e$-$02 \\
$n_{3}^{(i)}$ &   7.18333e$-$03 &   1.37046e$-$02 &  $-$2.22405e$-$02 &   1.97996e$-$02 &  $-$6.71756e$-$03 \\
$n_{4}^{(i)}$ &   3.10051e$-$03 &   3.88076e$-$03 &  $-$4.07139e$-$03 &   1.22677e$-$03 &  $-$3.58180e$-$04 \\
$d_{1}^{(i)}$ &   2.58979e$-$01 &   4.42756e$-$01 &   6.74009e$-$01 &   6.44566e$-$01 &  $-$4.79794e$-$01 \\
$d_{2}^{(i)}$ &   1.05613e+00 &   1.31787e+00 &   9.98757e$-$01 &   6.30569e$-$01 &   1.67564e$-$02 \\
$d_{3}^{(i)}$ &   4.35593e$-$01 &   6.79640e$-$01 &   5.08309e$-$01 &   2.20645e$-$01 &  $-$2.49549e$-$01 \\
$d_{4}^{(i)}$ &   1.68734e+00 &  $-$4.31052e$-$01 &   1.06194e$-$01 &   2.17586e$-$01 &  $-$2.29057e$-$01 \\
\hline
% Fit results for v1z_-1a1y with DBW2
& \cc{DBW2}\\ 
$c^{(i)}$ &   1.65353e$-$03 &  $-$5.20018e$-$03 &   5.58906e$-$03 &  $-$2.68645e$-$03 &   4.85235e$-$04 \\
$n_{1}^{(i)}$ &   3.05240e$-$02 &   1.73659e$-$02 &  $-$2.07260e$-$02 &   1.51523e$-$02 &  $-$5.59690e$-$03 \\
$n_{2}^{(i)}$ &  $-$3.78827e$-$02 &   1.51961e$-$02 &   1.31476e$-$02 &  $-$9.24480e$-$03 &   4.14817e$-$03 \\
$n_{3}^{(i)}$ &   3.28025e$-$02 &   6.55780e$-$03 &  $-$1.23069e$-$02 &   1.14891e$-$02 &  $-$6.25979e$-$03 \\
$n_{4}^{(i)}$ &  $-$1.50078e$-$03 &   1.08107e$-$02 &  $-$9.76440e$-$03 &   2.91846e$-$03 &  $-$2.97660e$-$04 \\
$d_{1}^{(i)}$ &   1.69736e+00 &   6.54862e$-$01 &   6.24044e$-$01 &   4.66231e$-$01 &  $-$4.83184e$-$01 \\
$d_{2}^{(i)}$ &   1.75714e+00 &   6.50739e$-$01 &   4.07680e$-$01 &   2.75134e$-$01 &  $-$7.55484e$-$02 \\
$d_{3}^{(i)}$ &   8.31723e$-$02 &   1.05494e+00 &   8.70566e$-$01 &   3.58793e$-$01 &  $-$4.81543e$-$01 \\
$d_{4}^{(i)}$ &   2.94855e+00 &  $-$1.89555e$-$01 &   2.11226e$-$02 &  $-$1.18785e$-$01 &  $-$1.54914e$-$01 \\
\hline

   \hline
 \end{tabular}
 \caption{Numerical results of the fit parameters for $c_{V_k}^-$.}
 \label{tab:v1z}
\end{table}
\begin{table}[htb]
 \begin{tabular}{c|rrrrr}
   $i$&0&1&2&3&4\\
   \hline
   % Fit results for v4z_-1a4y with plaquette
& \cc{plaquette}\\ 
$c^{(i)}$ &   3.50525e$-$03 &  $-$4.68207e$-$02 &   8.77916e$-$02 &  $-$6.22507e$-$02 &   1.48169e$-$02 \\
$n_{1}^{(i)}$ &  $-$3.76844e$-$01 &   2.12957e+00 &  $-$4.14093e+00 &   3.00354e+00 &  $-$7.25757e$-$01 \\
$n_{2}^{(i)}$ &  $-$2.06844e$-$01 &   7.70017e$-$01 &  $-$9.00837e$-$01 &   4.25447e$-$01 &  $-$7.25260e$-$02 \\
$n_{3}^{(i)}$ &  $-$2.78087e$-$01 &   2.41752e+00 &  $-$4.31448e+00 &   2.94944e+00 &  $-$6.34683e$-$01 \\
$n_{4}^{(i)}$ &  $-$1.82981e$-$01 &   2.78762e$-$01 &  $-$4.96115e$-$01 &   3.48106e$-$01 &  $-$9.33642e$-$02 \\
$d_{1}^{(i)}$ &   3.68004e+01 &   7.08098e+00 &  $-$3.92992e$-$02 &  $-$5.20857e+00 &   1.47375e+00 \\
$d_{2}^{(i)}$ &   1.30394e+01 &   2.67925e+01 &  $-$6.61332e+00 &  $-$2.84779e+01 &   1.77152e+01 \\
$d_{3}^{(i)}$ &   3.07591e+01 &   2.40240e+01 &  $-$1.04165e+00 &  $-$1.79010e+01 &   1.01610e+01 \\
$d_{4}^{(i)}$ &   5.72652e+00 &   1.12568e+01 &  $-$1.88964e+01 &   1.17469e+01 &  $-$1.68478e+00 \\
\hline
% Fit results for v4z_-1a4y with Iwasaki
& \cc{Iwasaki}\\ 
$c^{(i)}$ &   1.53487e$-$03 &  $-$3.41468e$-$02 &   6.82220e$-$02 &  $-$5.09028e$-$02 &   1.27728e$-$02 \\
$n_{1}^{(i)}$ &  $-$8.27139e$-$02 &   1.34471e+00 &  $-$2.86275e+00 &   2.20771e+00 &  $-$5.68265e$-$01 \\
$n_{2}^{(i)}$ &   1.03514e$-$01 &   1.49986e+00 &  $-$2.05986e+00 &   1.13542e+00 &  $-$2.11033e$-$01 \\
$n_{3}^{(i)}$ &   1.31475e$-$02 &   2.05314e+00 &  $-$3.25439e+00 &   2.04700e+00 &  $-$4.10781e$-$01 \\
$n_{4}^{(i)}$ &  $-$4.37623e$-$02 &   2.63972e$-$01 &  $-$4.70760e$-$01 &   3.26637e$-$01 &  $-$7.59031e$-$02 \\
$d_{1}^{(i)}$ &   3.30552e+01 &   4.09568e+00 &  $-$5.98710e+00 &   6.09342e+00 &  $-$2.34039e+00 \\
$d_{2}^{(i)}$ &   1.77638e+01 &   4.26837e+01 &  $-$6.91271e+00 &  $-$3.91159e+01 &   2.02790e+01 \\
$d_{3}^{(i)}$ &   3.03378e+01 &   2.33940e+01 &  $-$4.55782e+00 &  $-$1.45824e+01 &   7.89310e+00 \\
$d_{4}^{(i)}$ &   3.68914e+00 &   8.68953e+00 &  $-$1.06404e+01 &   3.38207e+00 &   6.35309e$-$01 \\
\hline
% Fit results for v4z_-1a4y with DBW2
& \cc{DBW2}\\ 
$c^{(i)}$ &   2.37474e$-$03 &  $-$1.26844e$-$02 &   2.50575e$-$02 &  $-$2.07670e$-$02 &   5.76665e$-$03 \\
$n_{1}^{(i)}$ &   6.48432e$-$02 &   1.36343e$-$01 &  $-$3.45284e$-$01 &   3.37077e$-$01 &  $-$1.04308e$-$01 \\
$n_{2}^{(i)}$ &   5.92327e$-$02 &   1.00062e$-$01 &  $-$5.62452e$-$02 &   2.12727e$-$02 &   1.20138e$-$02 \\
$n_{3}^{(i)}$ &   1.04664e$-$01 &   4.89947e$-$01 &  $-$4.13500e$-$01 &   1.42695e$-$01 &   2.24828e$-$03 \\
$n_{4}^{(i)}$ &  $-$1.51902e$-$03 &   1.41416e$-$01 &  $-$6.39386e$-$02 &  $-$1.09949e$-$01 &   6.86365e$-$02 \\
$d_{1}^{(i)}$ &   4.96718e+00 &  $-$1.30295e+00 &  $-$1.39159e+00 &   6.26564e+00 &  $-$2.70725e+00 \\
$d_{2}^{(i)}$ &   2.70372e+00 &   8.98447e+00 &  $-$1.36186e+00 &  $-$8.09935e+00 &   4.70601e+00 \\
$d_{3}^{(i)}$ &   6.12393e+00 &   8.34213e+00 &  $-$5.83018e$-$01 &  $-$3.18981e+00 &   1.53868e+00 \\
$d_{4}^{(i)}$ &   1.28638e$-$01 &   6.27763e+00 &  $-$3.24311e+00 &  $-$4.43441e+00 &   2.88724e+00 \\
\hline

   \hline
 \end{tabular}
 \caption{Numerical results of the fit parameters for $c_{V_0}^-$.}
 \label{tab:v4z}
\end{table}
\begin{table}[htb]
 \begin{tabular}{c|rrrrr}
   $i$&0&1&2&3&4\\
   \hline
   % Fit results for v1r_1a1r with plaquette
& \cc{plaquette}\\ 
$c^{(i)}$ &   0.00000e+00 &   0.00000e+00 &   0.00000e+00 &   0.00000e+00 &   0.00000e+00 \\
$n_{1}^{(i)}$ &   2.09397e$-$03 &  $-$5.33186e$-$04 &  $-$4.65192e$-$03 &   5.42610e$-$03 &  $-$2.24704e$-$03 \\
$n_{2}^{(i)}$ &  $-$1.06471e$-$02 &   2.27643e$-$02 &   4.01977e$-$02 &  $-$6.10584e$-$02 &   2.45592e$-$02 \\
$n_{3}^{(i)}$ &   1.38706e$-$02 &   5.94564e$-$02 &   1.64618e$-$02 &   4.22552e$-$02 &  $-$3.48815e$-$02 \\
$n_{4}^{(i)}$ &  $-$1.08445e$-$03 &  $-$1.93452e$-$03 &   7.00296e$-$03 &  $-$8.09199e$-$03 &   2.98526e$-$03 \\
$d_{1}^{(i)}$ &   1.22751e+00 &   2.94194e+01 &   1.07318e+01 &  $-$5.53447e+00 &  $-$1.88793e+00 \\
$d_{2}^{(i)}$ &  $-$9.43352e+00 &   1.44250e+01 &   7.73519e$-$01 &  $-$6.89522e+00 &   3.22929e+00 \\
$d_{3}^{(i)}$ &   1.12565e+01 &   1.83971e+01 &   1.66376e+01 &  $-$1.04853e+00 &  $-$7.23085e+00 \\
$d_{4}^{(i)}$ &  $-$1.75364e$-$01 &   2.06473e+00 &   8.42623e+00 &  $-$7.65891e+00 &   1.65849e+00 \\
\hline
% Fit results for v1r_1a1r with Iwasaki
& \cc{Iwasaki}\\ 
$c^{(i)}$ &   0.00000e+00 &   0.00000e+00 &   0.00000e+00 &   0.00000e+00 &   0.00000e+00 \\
$n_{1}^{(i)}$ &   8.71136e$-$03 &   1.34446e$-$02 &  $-$6.72308e$-$02 &   6.01238e$-$02 &  $-$1.63627e$-$02 \\
$n_{2}^{(i)}$ &  $-$4.80524e$-$02 &   9.21777e$-$02 &   5.29502e$-$01 &  $-$6.09020e$-$01 &   1.71269e$-$01 \\
$n_{3}^{(i)}$ &   5.71428e$-$02 &   5.68763e$-$01 &  $-$3.91779e$-$01 &   2.55005e$-$01 &  $-$1.05600e$-$01 \\
$n_{4}^{(i)}$ &  $-$6.84520e$-$03 &  $-$7.56285e$-$03 &   1.02776e$-$02 &  $-$1.11000e$-$02 &   5.96372e$-$03 \\
$d_{1}^{(i)}$ &   2.36201e+01 &   1.85591e+02 &   1.27770e+02 &  $-$2.20287e+02 &   5.36880e+01 \\
$d_{2}^{(i)}$ &  $-$7.63923e+01 &   8.29550e+01 &  $-$7.43883e+01 &   5.61588e+01 &  $-$9.22475e+00 \\
$d_{3}^{(i)}$ &   6.70337e+01 &   2.16961e+02 &  $-$4.95764e+01 &  $-$4.63254e+01 &   1.00042e+00 \\
$d_{4}^{(i)}$ &  $-$4.14600e+00 &   4.16800e+01 &  $-$4.09622e+01 &   2.30321e+01 &  $-$5.12005e+00 \\
\hline
% Fit results for v1r_1a1r with DBW2
& \cc{DBW2}\\ 
$c^{(i)}$ &   0.00000e+00 &   0.00000e+00 &   0.00000e+00 &   0.00000e+00 &   0.00000e+00 \\
$n_{1}^{(i)}$ &   1.24968e$-$02 &   9.44758e$-$03 &  $-$3.42722e$-$02 &   3.14005e$-$02 &  $-$9.98443e$-$03 \\
$n_{2}^{(i)}$ &  $-$3.82921e$-$02 &   1.39231e$-$01 &   4.08137e$-$01 &  $-$5.22501e$-$01 &   1.54012e$-$01 \\
$n_{3}^{(i)}$ &   4.77557e$-$02 &   4.30929e$-$01 &  $-$3.25076e$-$01 &   2.27980e$-$01 &  $-$9.17644e$-$02 \\
$n_{4}^{(i)}$ &  $-$5.77412e$-$03 &  $-$1.55592e$-$02 &   1.13224e$-$02 &  $-$8.06948e$-$03 &   4.26375e$-$03 \\
$d_{1}^{(i)}$ &   3.83004e+01 &   1.21779e+02 &   1.19599e+02 &  $-$1.84372e+02 &   4.54188e+01 \\
$d_{2}^{(i)}$ &  $-$9.12407e+01 &   5.90943e+01 &  $-$7.77565e+01 &   5.45309e+01 &  $-$4.40111e+00 \\
$d_{3}^{(i)}$ &   6.95720e+01 &   2.16484e+02 &  $-$4.76922e+01 &  $-$4.19054e+01 &  $-$8.56610e$-$01 \\
$d_{4}^{(i)}$ &  $-$3.56676e+00 &   2.24451e+01 &  $-$2.61850e+01 &   1.87052e+01 &  $-$4.82086e+00 \\
\hline

   \hline
 \end{tabular}
 \caption{Numerical results of the fit parameters for $c_{V_k}^H$.}
 \label{tab:v1r}
\end{table}
\begin{table}[htb]
 \begin{tabular}{c|rrrrr}
   $i$&0&1&2&3&4\\
   \hline
   % Fit results for v1s_1a1s with plaquette
& \cc{plaquette}\\ 
$c^{(i)}$ &   0.00000e+00 &   0.00000e+00 &   0.00000e+00 &   0.00000e+00 &   0.00000e+00 \\
$n_{1}^{(i)}$ &  $-$4.92854e$-$02 &   7.76273e$-$02 &  $-$1.09117e$-$01 &   5.41335e$-$02 &  $-$6.65490e$-$03 \\
$n_{2}^{(i)}$ &   1.62369e$-$02 &  $-$2.05156e$-$01 &   4.87357e$-$02 &  $-$1.00512e$-$02 &   2.19379e$-$02 \\
$n_{3}^{(i)}$ &  $-$1.62807e$-$02 &  $-$6.00090e$-$02 &   9.26995e$-$03 &  $-$9.52634e$-$02 &   5.14551e$-$02 \\
$n_{4}^{(i)}$ &  $-$1.34504e$-$02 &  $-$1.09913e$-$01 &  $-$9.93819e$-$03 &   3.72108e$-$02 &  $-$1.88461e$-$03 \\
$d_{1}^{(i)}$ &   1.07320e+00 &   8.61247e+00 &  $-$1.70831e+00 &   2.57570e+00 &  $-$2.35960e+00 \\
$d_{2}^{(i)}$ &  $-$6.07917e$-$01 &   7.62319e+00 &   4.48235e+00 &   1.52722e+00 &  $-$2.26997e+00 \\
$d_{3}^{(i)}$ &   8.80495e$-$01 &   8.95034e+00 &   1.54057e+00 &  $-$8.81352e$-$02 &  $-$1.36898e+00 \\
$d_{4}^{(i)}$ &   2.21861e$-$01 &   2.61262e+00 &   5.61100e$-$01 &  $-$1.25562e+00 &   1.61500e$-$01 \\
\hline
% Fit results for v1s_1a1s with Iwasaki
& \cc{Iwasaki}\\ 
$c^{(i)}$ &   0.00000e+00 &   0.00000e+00 &   0.00000e+00 &   0.00000e+00 &   0.00000e+00 \\
$n_{1}^{(i)}$ &  $-$1.89216e$-$02 &   7.61233e$-$02 &  $-$9.73709e$-$02 &   5.06130e$-$02 &  $-$8.17803e$-$03 \\
$n_{2}^{(i)}$ &  $-$7.31825e$-$03 &  $-$1.74543e$-$01 &   2.94655e$-$01 &  $-$1.91881e$-$01 &   5.14398e$-$02 \\
$n_{3}^{(i)}$ &  $-$7.05824e$-$02 &   2.71060e$-$01 &  $-$1.02481e$-$01 &   1.73657e$-$01 &  $-$6.37448e$-$02 \\
$n_{4}^{(i)}$ &  $-$9.00773e$-$03 &  $-$1.67444e$-$01 &   1.36274e$-$01 &  $-$1.33701e$-$01 &   5.65792e$-$02 \\
$d_{1}^{(i)}$ &  $-$1.97613e+00 &   3.89137e+01 &   1.57804e+00 &   2.54447e+00 &  $-$6.67567e+00 \\
$d_{2}^{(i)}$ &   4.57630e+00 &   3.62223e+01 &   2.26543e+01 &   9.04292e+00 &  $-$1.38239e+01 \\
$d_{3}^{(i)}$ &   3.86975e+00 &   3.56956e+01 &   3.56635e+01 &   1.22621e+01 &  $-$1.85228e+01 \\
$d_{4}^{(i)}$ &   8.34828e$-$02 &   1.73013e+01 &   7.67351e+00 &  $-$3.68814e+00 &  $-$1.92297e+00 \\
\hline
% Fit results for v1s_1a1s with DBW2
& \cc{DBW2}\\ 
$c^{(i)}$ &   0.00000e+00 &   0.00000e+00 &   0.00000e+00 &   0.00000e+00 &   0.00000e+00 \\
$n_{1}^{(i)}$ &   3.94399e$-$02 &   6.12762e$-$02 &  $-$4.42999e$-$02 &   2.91012e$-$03 &   5.62763e$-$03 \\
$n_{2}^{(i)}$ &  $-$5.50451e$-$04 &   1.53894e$-$02 &   7.33107e$-$02 &  $-$2.14443e$-$02 &   1.26660e$-$02 \\
$n_{3}^{(i)}$ &   1.44008e$-$02 &  $-$5.31159e$-$03 &   5.91542e$-$02 &  $-$7.08047e$-$02 &   3.46751e$-$02 \\
$n_{4}^{(i)}$ &   2.32681e$-$02 &   8.14410e$-$02 &  $-$3.26852e$-$02 &   6.02456e$-$02 &  $-$3.77997e$-$02 \\
$d_{1}^{(i)}$ &   1.45225e+00 &   1.06854e+00 &   8.00493e$-$01 &   4.90831e$-$01 &  $-$1.84695e$-$01 \\
$d_{2}^{(i)}$ &   4.10400e$-$01 &   6.60514e$-$01 &   6.29942e$-$01 &   5.51064e$-$01 &   2.32655e$-$01 \\
$d_{3}^{(i)}$ &   1.24019e+00 &   1.09508e+00 &   7.99430e$-$01 &   4.32776e$-$01 &  $-$4.18890e$-$01 \\
$d_{4}^{(i)}$ &   8.12694e$-$01 &   1.16732e+00 &   7.68715e$-$01 &   2.41916e$-$01 &  $-$6.06008e$-$01 \\
\hline

   \hline
 \end{tabular}
 \caption{Numerical results of the fit parameters for $c_{V_k}^L$.}
 \label{tab:v1s}
\end{table}

 \clearpage
\clearpage
\clearpage
\clearpage
\clearpage
\begin{figure}[p]
 \begin{center}
  \includegraphics*[width=11cm,clip=true]
  {plaq_cvkHL.eps}\\
  \includegraphics*[width=11cm,clip=true]
  {iwasaki_cvkHL.eps}\\
  \includegraphics*[width=11cm,clip=true]
  {dbw2_cvkHL.eps}
 \end{center}
 \caption{The same figure as Fig.~\ref{fig:v1x} but for $c_{V_k}^H$
 (circles) and $c_{V_k}^L$ (triangles).}
 \label{fig:v1rs}
\end{figure}
\clearpage

\end{document}